\documentclass{aa}
\usepackage{rotating}
\usepackage[utf8]{inputenc}
\usepackage{amssymb}
\usepackage{graphicx}

\usepackage{lscape}
\usepackage{float}
\usepackage{txfonts}
\usepackage{hyperref}
\usepackage{afterpage}
\usepackage{tablefootnote}
\usepackage{threeparttable}
\title{Eccentric binaries: Periastron events and tidal heating}

   \author{ G. Koenigsberger\thanks{Visiting scholar, Astronomy Department, Indiana University}
            \inst{1}
        \and
            D. Estrella-Trujillo
           \inst{1}
          }

   \institute{ Instituto de Ciencias F\'{\i}sicas, Universidad Nacional Aut\'onoma de M\'exico,
              Ave. Universidad S/N, Chamilpa, Cuernavaca, M\'exico \\
              \email{gloria@icf.unam.mx, dtrujillo@icf.unam.mx}
             }

\date{August 2023}

\abstract
{Binary stars cannot be fully understood without assessing the interaction effects between the two
components and the impact of these effects on observational diagnostics. Periastron brightening events,
also known as the heartbeat phenomenon, are a clear manifestation of this type of interaction.
}
{We aim to explore the  role of tidal shear energy dissipation in stars undergoing periastron 
brightening events.
}
{We performed a computation from first principles that uses  a quasi-hydrodynamic Lagrangian scheme 
to simultaneously solve the orbital motion  and the equations of motion of a 3D grid of volume elements
covering the inner, rigidly rotating "core"  of a rotating, the tidally perturbed star.  The equations 
of motion include the gravitational acceleration of both stars, the centrifugal, Coriolis, gas pressure 
accelerations, and viscous coupling between volume elements.  The method is illustrated for a grid of
model binary systems with a 10 M$_\odot$ primary that is perturbed by a 6.97 M$_\odot$ companion in
eccentric orbits ($e$=0 - 0.7).  The model was then applied to the heartbeat star MACHO 80.7443.1718.
}
{We find an increase by factors $\sim$10$^{-6}$ -10$^{-3}$ in tidal shear energy dissipation at periastron,
consistent with the majority of observed heartbeat stars. The magnitude of the periastron effect
correlates with the degree of departure from synchronicity:  stars rotating much faster or much slower 
than the synchronous rate at periastron present the strongest effect. We confirm that for eccentricities 
$\leq$0.3, pseudo-synchronization occurs for 0.8$<\omega/\Omega_{ave}<$1, where $\Omega_{ave}$ is the average 
orbital angular velocity.  The  minimum energy rotation rate (pseudo-synchronism) for $e$=0.5 and 0.7  
occurs for 1.0$<\omega/\Omega_{ave}<$1.15.  The tidal shear energy dissipation model reproduces 
from first principles the $\sim$23\% maximum brightness enhancement at periastron of MACHO 80.7443.1718. 
}
{Our results suggest that the magnitude and shape of the heartbeat signal may serve as diagnostics for
the internal stellar rotation and turbulent viscosity values. 
}
\newcommand\vv{{\mathrm v}  }
\begin{document}

\maketitle

\section{Introduction}

A growing number of binary systems have been reported to undergo periastron events, that is,
a change in the observable properties around the time of periastron,  such as a 
brightness enhancement, an increase in surface activity, and/or a triggering of oscillations.
Early examples include \citet{1979IAUS...83....3H}, who cited a         
significant increase in mass-loss rates around periastron in the two B-type systems, HD 187399 and
AZ Cas, and in HD 163181 (B0Ia+0), and \citet{2007IBVS.5782....1V}
at periastron in six systems, including $\eta$ Carinae.


More recently, photometric observations obtained with the {\it Kepler} space telescope have led 
to the discovery of objects that not only display periastron-related brightness enhancements, but 
also low-amplitude sinusoidal oscillations that tend to have frequencies that are harmonics of the 
orbital frequency \citep{2012ApJ...753...86T, 2017MNRAS.472.1538F, 2020svos.conf..203G}.   These objects 
have come to be known as heartbeat stars, and more than 180 are known from Kepler alone 
\citep{2013EAS....64..285H, 2016AJ....151...68K} and many more from BRITE \citep{2018pas7.conf..151P}, 
ASAS-SN \citep{2020MNRAS.491...13J}, and TESS  \citep{2021A&A...647A..12K}.  These stars are 
all considered to be eccentric orbit binary systems with typical photometric 
variations  $\Delta$L/L$\leq$10$^{-3}$.


Although a large majority of the heartbeat stars have masses in the range of A-type and F-type stars,
similar properties are also reported in  high-mass OB-type objects such as $\iota$ Orionis 
\citep{2017MNRAS.467.2494P} and $\eta$ Carinae \citep{2018MNRAS.475.5417R} in the Milky Way, 
MACHO  80.7443.1718   \citep{2019MNRAS.489.4705J} in the Large Magellanic Cloud, and the
massive multiple system HD\,5980 \citep{1997A&A...328..269S, 2021A&A...647A..12K}
in the Small Magellanic Cloud.  This large diversity  points to a common physical mechanism which 
is now widely accepted to be the variable gravitational perturbation caused by the binary 
companion.\footnote{It must be noted that periastron brightening in 
massive stars is often interpreted in terms of colliding winds, a phenomenon that is however 
hard-pressed to explain low-amplitude high-frequency post-periastron oscillations.} 

A companion distorts a star's intrinsic spherical shape producing
two tidal lobes, one facing the companion and one in the opposite hemisphere.  The extension of
the tidal lobes is not necessarily the same on both sides and their orientation is not necessarily 
parallel to the line joining the stars' centers.  These depend on the orbital separation as well as
the star's radius, its rotation velocity, and the viscosity of its material. 

In binary systems with circular orbits, the shape of the distorted stars remains approximately constant 
throughout the orbital cycle. However, because the viewing angle to the deformed star changes with 
orbital phase, the changing projected size of the disk in the sky leads to brightness variations. 
This so-called ellipsoidal effect is a purely geometric effect, and can be modeled using an 
appropriate stellar atmosphere model \citep{2000A&A...364..265O}.

Eccentric binaries differ from circular orbit systems in that the orientation of the tidal bulges 
with respect to the line joining the two stellar centers varies over the orbital cycle, as does
their extension. These variations occur most rapidly around periastron, triggering high
frequency dynamical effects that show up as low-amplitude photometric variations.  

\citet{1977ApJ...213..183P} presented a mathematical framework to model the tidal distortions 
based on a linear superposition of the non-radial pressure and gravity mode oscillations, a framework 
further developed by \citet{1995ApJ...449..294K} and \citet{1997ApJ...490..847L}, predicting the brightening 
around periastron and the appearance of tidally excited oscillations (TEOs) that are harmonics of 
the orbital period.  
Recent advances were achieved by \citet{2017MNRAS.472.1538F} to account for rotation, incorporate 
nonadiabatic effects, and allow for spin-orbit misalignment; \citet{2001PhRvD..65b4001S} who
studied the nonlinear interaction of modes in rotating Newtonian stars; and by \citet{2024MNRAS.527.2288D} who 
proposed an alternative approach to the treatment of tidally driven oscillations damping that accounts for 
dissipative mode coupling. 


As with any theoretical framework, the linear mode-decomposition models have limitations.  As pointed 
out by \citet{1977ApJ...213..183P}, a linear mode calculation is valid if nonlinear effects are 
very small.  Specifically, with $R_1$ the stellar radius and $r_{per}$ the periastron distance, 
they point out that the validity of a linear mode calculation is to be doubted if 
$(R_1/r_{per})^2\geq0.10$. Indeed, using a 
3D non-perturbative modeling method, \citet{2024arXiv240102573F} show that the linear approach 
underestimates tidal interactions by up to 40\% for binary stars with small separations.   
In addition, the tidal interaction imposes a large-scale differential-rotation structure  as 
well as induced oscillatory shearing motions on orbital timescales \citep{2021A&A...653A.127K}.  
An important consequence of the differential rotation is that the shearing motions can be 
damped through the presence of turbulent viscosity, thus converting kinetic energy into heat 
in a process known as tidal shear energy dissipation.   However, although observational 
evidence exists for deviations from an adiabatic approximation 
\citep{2020ApJ...888...95G}, with a few exceptions (see, for example, 
\citet{2013JFM...719...47B, 2021A&A...647A.144A}) the current analytical models assume a 
rigidly rotating star and thus neglect tidal shear energy heating. Injection of this added 
heat into a star's subphotospheric layers causes it to expand and become more luminous, at 
least in the case of a solar-type star \citep{2023A&A...670A..44E}.  



In the presence of turbulent viscosity, kinetic energy in the tidal flows is converted into heat at 
all orbital phases, not only at periastron.  However, as the star approaches and recedes from periastron, 
the shearing layers undergo rapid velocity variations leading to a more abrupt release of energy. 
\citet{Moreno:2011jq}  investigated the potential role of tidal heating as a source of periastron 
brightening events, concluding that it could play a significant role in this phenomenon. However, 
they were unable to provide quantitative constraints for three reasons. The first is that the study was 
limited to a single surface layer, thus neglecting the effects of underlying layers.  The second 
is that the model needed as input a turbulent viscosity, values of which are highly uncertain. The
third reason is that  the viscosity remained constant throughout the calculation, whereas in reality 
turbulent viscosity is non-isotropic and time variable. These three limitations have now been overcome 
\citep{2021A&A...653A.127K, 2023A&A...670A..44E} enabling this study.

The objective of this paper is to show the general features associated with tidal shear energy dissipation 
in eccentric binary systems, particularly the increased brightness predicted to occur around the time
of periastron.  The method is summarized in Section 2. In Section 3 we present the results from a grid
of model calculations for a grid of binary systems having a range of orbital eccentricities and
rotation velocities.  In Section 4 we apply the model to one of the most extreme heartbeat stars found 
to date, MACHO 80.7443.1718, and illustrate the manner in which our calculation from first principles
directly predicts the observed variability amplitude.  A discussion and the conclusions are 
presented in Section 5.

\section{Method}

Binary stars are defined to be in equilibrium when the orbit is circular, the rotation axis coincides
with the orbital angular velocity vector, and the rotation period is equal to the orbital period.
A departure from any of these conditions induces deformations of the stellar surface and excites
oscillations.

Classical tidal theory divides the analytical treatment of the perturbations into the equilibrium tide model
(Darwin 1880; Alexander 1973; Hut 1981, 1982) and the dynamical tide model
\citep{1975A&A....41..329Z, 1987ApJ...318..261M, 1989ApJ...342.1079G, 1997ApJ...490..847L,
1995MNRAS.277..471S}.
The first analyzes the large-scale distortion imposed by the external gravitational perturber.  It
assumes that the viscosity of the material leads to a lag between the principal axis of the deformed
spheroid and the line of centers of the two stars.  This lag is responsible for the torque that over time
synchronizes the rotation and circularizes the orbit, and it is caused by the fact that the stellar material
is not inviscid.  The second model treats the star as an oscillator
with a number of modes that are excited by the gravitational potential of the companion, and assumes
that the dissipation of these oscillations is responsible for the eventual synchronization and
circularization of the orbit (Ogilvie 2014 and references therein). As noted by \citet{1998ApJ...499..853E},
rather than competing, each model provides a useful description of the interaction under different
regimes; that is, for different masses and radii, orbital separations and eccentricities, and rotation rates.
However, both models suffer from important uncertainties in the treatment of dissipation
\citep{2014ARA&A..52..171O,2016A&A...592A..33M, 2021MNRAS.506L..69B, 2021MNRAS.503.5789T,
2023ApJ...953...48T, 2023ApJ...945...43S}.  Both models are intended for studying the long term
evolution of the stellar rotation and orbital properties of binary stars, with a focus on
determining the timescales over which an equilibrium  configuration is reached.  However, the uncertainty
in the dissipation limits the possibility of quantitative results.

The approach used in this paper differs significantly from the two classical models. It makes no
a priori assumptions other than that the damping mechanism can be associated with the viscous
stresses suffered by fluid parcels that are in relative motion with respect to each other.  Such
relative motions naturally occur throughout the star given the non-isotropic dependence of the
tidal forcing in the radial and horizontal directions. Thus, our method relies on a viscosity parameter
as it appears in the shear part of the kinematic stress tensor (e.g., \citet{1971Symon}). The
aim is to analyze the short term (orbital timescales) variations induced by the tidal interaction 
in order to allow a direct comparison with observational data. 

The method is a computation from first principles which is named TIDES\footnote{The {\it Tidal 
Interactions with Dissipation of Energy due to Shear} code in all its versions is available upon 
request to G. Koenigsberger and is easily implemented in any operating system running a Fortran 
or GNU Fortran compiler}.  It uses  a quasi-hydrodynamic Lagrangian scheme which  simultaneously solves 
the orbital motion of the companion and the equations of motion of a grid of volume elements
covering the inner, rigidly rotating "core"  of the tidally perturbed primary star.  The core is
defined as the interior region that is rotating as a solid body and does not necessarily coincide
with the nuclear burning region.  The equations of motion include the gravitational acceleration
of both stars as well as centrifugal, Coriolis, and gas pressure accelerations.  The motions of
individual elements are coupled to those of neighboring elements and to the core through viscous 
stresses.  

The numerical scheme was first applied to the B-type system $\iota$ Orionis \citep{1999RMxAA..35..157M}.
It predicted a rapid increase in the stellar radius around periastron as well as small-amplitude,
short-timescale oscillations throughout the orbit. It was subsequently used to produce synthetic
absorption line profiles as a function of the orbital phase in {\it $\epsilon$ Persei} \citep{Moreno:2005cq} 
and $\alpha$ Virginis (Harrington et al. 2009; 2014; Palate et al. 2013b) and other systems
\citep{2013A&A...552A..39P}.  The simulated line-profile
variability captures many of the characteristics of the observations, namely, the spectroscopic
equivalent of the ellipsoidal effects and the traveling ``bumps'' that are characteristic of
tidally induced non-radial pulsators. The method was also used to derive the  synchronization
timescales for eccentric binaries \citep{2007A&A...461.1057T, Moreno:2011jq}, finding  good
agreement with the expression given by \citet{2008EAS....29...67Z} (his Eq. 2.5)  for viscous dissipation
for  circular and $e$=0.1 orbits, the cited range of applicability.  

The above calculations were performed for a single surface layer contiguous to the rigidly rotating 
internal region.  The output includes the tidal perturbations in the radial and azimuthal directions 
and it was shown that the tidal velocity in the radial direction, $\vv_r$, is generally an order of 
magnitude smaller than that in the azimuthal direction, $\vv_\varphi$, consistent with the analytical 
results for the equatorial latitudes\footnote{The inviscid approximation leads to poorer results for the 
tidal amplitude at latitudes closer to the poles because it neglects angular momentum to these 
regions from the strongly perturbed equatorial latitudes, see \citet{2009ApJ...704..813H}, Fig. 7.}
of \citet{1981ApJ...246..292S}, derived by assuming small departures from corotation.  The dominance
of the azimuthal component of the tidal flows was also found by \citet{2017MNRAS.468.1864L}.  
This original version of the numerical method 
is named {\it TIDES-1}.  It requires a viscosity value as input and its output includes the 
radial and azimuthal velocities for the surface layer. These velocities on the hemisphere 
facing the observer are then projected onto the line of sight allowing the computation of the 
orbital phase-dependent profile variability of photospheric absorption line profiles \citep{Moreno:2005cq}.

The extension of the TIDES method to $n$ layers (TIDES-n) was introduced in Koenigsberger et al. (2021) 
where, for circular-orbits, it was shown that the rotation rate tends toward a uniform rotation structure
when synchronization is attained.  However, the synchronization timescale is different for each layer,
because it goes as the viscous timescale, $\sim\nu$/r$^2$, where $\nu$ is the viscosity and $r$ is 
the radius of the particular layer. The calculation in Koenigsberger et al. (2021) showed how each
layer procedes toward synchronization and that the average angular velocity of the surface layer 
(at the equator) is the first to reach synchronicity, in agreement with the conclusion reached 
analytically by \citet{1989ApJ...342.1079G}.  TIDES-n requires a turbulent viscosity value as 
input.  It is isotropic and remains constant throughout the calculation. 

The primary results presented in this paper use the latest version of the code (TIDES-nvv). It is
based on the same algorithm as TIDES-n but has the additional capability of computing a
turbulent viscosity at each grid point \citep{2023A&A...670A..44E}. It  outputs the azimuthal
velocity, turbulent viscosity and corresponding energy dissipation rate.  These variables are
all non-isotropic and time-variable.


The strongest approximation in our calculation concerns the  mathematical formulation for the
turbulent viscosity.  Turbulent viscosity arises in unstable flows that are characterized
as turbulent. \citet{2004A&A...425..243M} noted that turbulence might be expected even in 
stellar radiative zones due to nonlinear processes that appear at large Reynolds numbers.  
Understanding turbulent viscosity is a complex problem that falls in
the realm of fluid dynamics and turbulent transport (see, for example, \citet{2022SGeo...43..229L,
1992A&A...265..115Z, 1999A&A...347..734R, 2008EAS....29...67Z, 2004A&A...425..243M,
2016A&A...592A..33M, 2016ApJ...821...49G}). We use the simplest functional representation,
as given by \citep{1987flme.book.....L}, in which its value is directly proportional to the
instantaneous gradient in velocity of neighboring fluid parcels.  This representation  yields
a non-isotropic and time-dependent turbulent viscosity value, which is a significantly less
arbitrary approach than simply adopting a constant value.  Furthermore, the viscosity thus
derived includes both the  shear due to the global differential rotation and the shear
due to the  tidally excited oscillations.

The need for a turbulent viscosity in massive star models follows from the fact that rotation 
induced shear is required to account for the enhanced abundances of nuclear-processed elements on
the surface of main sequence massive stars \citep{1996A&A...313..140M, 1997A&A...322..209T,
2000ARA&A..38..143M}. The fluid velocity used in these formulations is associated with meridional
circulation currents and the differential effect of the Coriolis force \citep{2010A&A...519A..16H}.
In binary stars with asynchronous rotation, the tidal perturbations can induce significantly larger 
fluid velocities and velocity gradients (Koenigsberger et al. 2021).  Also, observational data of binary systems 
in clusters indicates that systems with orbital periods shorter than $\sim$6-8 days have mostly 
circular orbits, which means that the circularization process must occur on much shorter timescales 
than possible in the presence of only molecular viscosity.  The circularization timescale that is 
usually applied in this case is as derived by \citet{1977A&A....57..383Z}, with typical turbulent 
viscosities in the range 10$^{10}$cm$^2$ s$^{-1}$-10$^{13}$cm$^2$ s$^{-1}$  \citep{2008EAS....29...67Z}.
It is also found that turbulent viscosity as estimated using arguments from the mixing length
theory should have values on the order of 10$^{12}$-10$^{13}$ cm$^2$/s \citep{2020A&A...636A..93K}.

It is interesting to note that viscosity values that are orders of magnitude larger than the molecular 
(kinematical) viscosities (1-10$^3$ cm$^2$/s) appear in many additional contexts.  For example,
numerical simulations of the solar convective region use a viscosity value 10$^{10}$-10$^{12}$ cm$^2$/s        
\citep{2000ApJ...533..546E} based on the idea that the dynamics are expected to be governed principally
by the largest (and most energetic) eddies, and not to be sensitive to small-scale flow features.
\footnote{\citet{2000ApJ...533..546E} note, however, that the
large viscosity used in their models should be regarded as an artificial viscosity whose values are
chosen to suitably truncate the nonlinear energy cascade and thereby prevent nonlinear instability.}
This is in contrast to the value of kinematic viscosity  $\sim$27 cm$^2$/s at the solar center
discussed in Fig. 3 of \citet{2016MNRAS.460..338C}, where the plasma conditions are significantly
different from those in the convective zone.
Also, \citet{1999ApJ...525L..65P} estimated $\nu$=6. 10$^{13}$ cm$^2$/s for this parameter in boundary 
layers around solar system planetary ionospheres; \citet{2018Icar..300..223E} used the outgoing 
energy flux due to the vapor plumes in Saturn's moon Enceladus to estimate that its mean tidal 
viscosity  $\eta$=$\nu\rho \sim$0.24 10$^{14}$ Pa~s (which translates into $\nu\sim$2 10$^{14}$   cm$^2$/s
assuming a mass density $\rho\sim$0.9 for ice);  and \citet{2006ScChD..49..492S}  found
$\eta \sim$ 10$^{19}$ Pa-s for Earth's crustal medium at a depth of 32 km.



\subsection{Definition of parameters}

The star whose tidal perturbations are being considered is called the "primary", and its mass and 
unperturbed radius are denoted, respectively,  $m_1$ and $R_1$. The companion is assumed to be a point 
source and its mass is labeled $m_2$. The primary's initial rotation is uniform and the rotation axis 
is perpendicular to the orbital plane.  A number $N_r$ of stellar layers are allowed to respond to 
the forces in the system.  This leads to a departure from the initial uniform rotation structure as it 
evolves over time.

We characterize the initial rotation velocity in terms of the synchronicity parameter 
$\beta_{0}$=$\omega_0/\Omega_{0}$, where $\omega_0$ is the initial rotation angular velocity and 
$\Omega_0$ is the orbital  angular velocity at a reference point.  For eccentric orbits, periastron 
is a convenient reference point, so that $\beta_0$ can be cast 
in the form: $\beta_{0}$=0.02P$\frac{V_{rot}(1-e)^{3/2}}{R_{1}(1+e)^{1/2}}$.
Here, $P$ is the orbital period in days, $V_{rot}$ is the initial equatorial rotation velocity of the
uniformly rotating star (initial condition) in km/s, and $R_1$ is the stellar radius in solar units.

\begin{table}[h!]
\caption{Description of input parameters. \label{table_modelsmi}}
\centering
\begin{tabular}{ c c l }
\hline\hline
Parameter&Value&Description\\ 
\hline
\centering
P$_{orb}$          & (a) &Orbital period (days)\\
e                  & (a) &Orbital eccentricity\\
m$_{1}$            &10       &Primary mass (perturbed star) (M$_{\odot}$)\\
m$_{2}$            &6.97     &Secondary mass (point source) (M$_{\odot}$)\\
R$_{1}$            &5.48     &Undisturbed primary radius (R$_{\odot}$)\\
$\beta_0$          & (a)      &Initial  synchronicity parameter\\
$\lambda$          & 1       &Turbulent viscosity coefficient\\
IND                &3        &Polytropic index\\
$\Delta$R/R$_{1}$  &0.06     &Layer thickness\\
N$_{r}$            &10       &Number of layers\\
N$_{\varphi}$      &200      &Number of partitions in longitude\\
N$_{\theta}$       &20       &Number of partitions in latitude\\
Tol                &10$^{-7}$&Tolerance for Runge-Kutta integration\\
\hline
\hline
\end{tabular}
\tablefoot{(a) See Tables~\ref{table_overview},~\ref{table_circular}, and~\ref{table_eccentric}.
}

\end{table}

\begin{table}[h]
\centering
\caption{Overview of the model grid. \label{table_overview}}
\begin{tabular}{ l c l l l l }
\hline\hline
Set&   $e$ &P  &$a$            &N$_{cy}$ & N$_{phases}$    \\
\hline
 E0.0&0.0 &6.00  &35.7   &30, 50  &10    \\
 E0.1&0.1 &6.00  &35.7   &30, 50  &10     \\
 E0.3&0.3 &8.75  &45.9   &30, 50  &130    \\
 E0.5&0.5 &14.49 &64.3   &12, 20  &130   \\
 E0.7&0.7 &31.18 &107.2  & 5, 10  &169  \\
\hline
\hline
\end{tabular}
\tablefoot{$e$ is the orbital eccentricity, the orbital period P is given in days,
and the orbital semimajor axis $a$ is given in R$_\odot$.  N$_{cy}$ is the number of
orbital periods at which the results were sampled and N$_{phases}$ is the number of
orbital phases at each of the sampled cycles for which output was written. 
}
\end{table}

The rate of energy dissipation per unit volume is as given in \citet{Moreno:2011jq}:

\begin{equation} \dot{E}_{V} \simeq {\nu \rho} \left \{ \frac{4}{3} \left (
\frac{\partial \omega}{\partial {\varphi}} \right )^2 +
\left [ r^{2} \left ( \frac{\partial \omega}{\partial r} \right )^2 +
\left ( \frac{\partial \omega}{\partial \theta} \right )^2
\right ]{\sin}^2{\theta} \right \},
\label{dis1} \end{equation}

\noindent with $\nu$  the kinematical viscosity, $\rho$ the mass density, $\omega$  the 
angular velocity, and $r$, $\theta$, $\varphi$, respectively, the radius, latitude, and 
longitude coordinates.  The units of $\dot{E}_{V}$ are energy per unit time per unit volume. 

The viscosity is assumed to be a turbulent viscosity and is calculated at each timestep and 
volume element interface using the expression \citep{1987flme.book.....L}:

\begin{equation}
 \nu_\mathrm{turb} =\lambda \ell_\mathrm{t} \Delta u_\mathrm{t},         \label{eq_Landau_Lifshitz}
\end{equation}

\noindent where $\ell_\mathrm{t}$ is the characteristic length of the largest eddies that are associated 
with the turbulence, here assumed to be on the order of the layer thickness, and $\Delta u_\mathrm{t}$ 
is the typical average velocity variation of the flow over the length $\ell_\mathrm{t}$.  The variable 
$\lambda\in$[0,1] is a proportionality parameter that describes the fraction of kinetic energy transformed 
into thermal energy.   We used $\lambda$=1. Inspection of Eqs.~\ref{dis1} and \ref{eq_Landau_Lifshitz} 
and numerical experiments (Estrella-Trujillo et al. 2024, in preparation) indicate that the energy 
dissipation rate for different values scales approximately with the value of $\lambda$ and is thus 
straightforward to estimate for smaller values.  The above prescription is applied to the interface 
between each volume element and its neighboring elements. Thus, the non-isotropic and time-variable 
nature of the viscosity is incorporated in the calculation.


The density is obtained assuming a polytropic stellar structure.  Each volume element of the grid
contains a fixed mass of gas that obeys the ideal gas equation of state, which provides the
corresponding gas pressure.

The total energy dissipation rate at particular times during the orbital cycle is:

\begin{equation}
\dot{E}_{tot}(t) = \int_V \dot{E}_V  dV.
\end{equation}

Another quantity that is of interest, especially when comparing with investigations of tidal theory, is
the energy dissipation rate averaged over the orbital cycle:

\begin{equation}
\left<\dot{E}_{tot}\right> = \frac{\int_0^P \dot{E}_{tot}(t) dt}{P}.
\end{equation}

\begin{table*}[h]
\centering
\caption{Eccentric models.  \label{table_eccentric}}
\begin{tabular}{ l l l l l l l }
\hline\hline
Num&  $\beta_0$ & V$_{rot}$ & $<\dot{E}_{tot}>$  & $\dot{E}_{tot}^{per}$ & $\dot{E}_{tot}^{ap}$ &$\Delta \dot{E}$/L$_*$  \\
\hline
\multicolumn{6}{c}{P=6.0 ~~~~~ $e$=0.1 ~~~~~ $r_{per}$=32.15} \\
\hline
 201&  0.20&    11&  7.60 10$^{-1}$ &  1.65           &  4.10 10$^{-1}$ &  3.10 10$^{-3}$\\
 202&  0.40&    22&  2.30 10$^{-1}$ &  5.40 10$^{-1}$ &  1.20 10$^{-1}$ &  1.05 10$^{-3}$\\
 203&  0.60&    34&  4.80 10$^{-2}$ &  1.40 10$^{-1}$ &  2.00 10$^{-2}$ &  3.00 10$^{-4}$\\
 204&  0.80&    45&  4.20 10$^{-3}$ &  1.20 10$^{-2}$ &  1.20 10$^{-3}$ &  2.70 10$^{-5}$\\
 205&  0.88&    49&  2.50 10$^{-3}$ &  4.80 10$^{-3}$ &  2.00 10$^{-3}$ &  7.00 10$^{-6}$\\
 206&  0.95&    53&  4.20 10$^{-3}$ &  2.40 10$^{-3}$ &  5.50 10$^{-3}$ & -7.75 10$^{-6}$\\
 207&  1.00&    56&  8.80 10$^{-3}$ &  3.40 10$^{-3}$ &  1.10 10$^{-2}$ & -1.90 10$^{-5}$\\
 208&  1.05&    59&  1.50 10$^{-2}$ &  7.10 10$^{-3}$ &  2.00 10$^{-2}$ & -3.22 10$^{-5}$\\
 209&  1.20&    67&  6.30 10$^{-2}$ &  4.90 10$^{-2}$ &  7.60 10$^{-2}$ & -6.75 10$^{-5}$\\
 210&  1.40&    79&  2.70 10$^{-1}$ &  2.70 10$^{-1}$ &  3.00 10$^{-1}$ & -7.50 10$^{-5}$\\
 211&  1.60&    90&  9.30 10$^{-1}$ &  9.30 10$^{-1}$ &  8.80 10$^{-1}$ &  1.25 10$^{-4}$\\
 212&  1.88&   105&  3.01           &  3.48           &  3.12           &  9.00 10$^{-4}$\\
 213&  1.95&   109&  4.20           &  4.84           &  4.32           &  1.30 10$^{-3}$\\
\hline
\multicolumn{6}{c}{P=8.73~~~~~ $e$=0.3 ~~~~~ $r_{per}$=32.1} \\
\hline
 301&  0.20&    12&  6.50 10$^{-1}$ &  2.47           &  1.80 10$^{-1}$ &  5.73 10$^{-3}$\\
 302&  0.40&    24&  1.80 10$^{-1}$ &  6.90 10$^{-1}$ &  3.50 10$^{-2}$ &  1.64 10$^{-3}$\\
 303&  0.60&    37&  4.90 10$^{-2}$ &  2.30 10$^{-1}$ &  1.00 10$^{-2}$ &  5.50 10$^{-4}$\\
 304&  0.80&    49&  2.40 10$^{-2}$ &  7.90 10$^{-2}$ &  1.10 10$^{-2}$ &  1.70 10$^{-4}$\\
 305&  0.88&    54&  1.60 10$^{-2}$ &  1.70 10$^{-2}$ &  9.60 10$^{-3}$ &  1.85 10$^{-5}$\\
 306&  0.95&    58&  1.30 10$^{-4}$ &  7.60 10$^{-3}$ &  1.30 10$^{-2}$ & -1.35 10$^{-5}$\\
 307&  1.00&    61&  2.20 10$^{-2}$ &  6.70 10$^{-3}$ &  1.60 10$^{-2}$ & -2.33 10$^{-5}$\\
 308&  1.05&    64&  2.40 10$^{-2}$ &  5.60 10$^{-3}$ &  1.70 10$^{-2}$ & -2.85 10$^{-5}$\\
 309&  1.20&    73&  5.30 10$^{-2}$ &  3.00 10$^{-2}$ &  3.60 10$^{-2}$ & -1.50 10$^{-5}$\\
 310&  1.40&    86&  1.70 10$^{-1}$ &  2.10 10$^{-1}$ &  1.20 10$^{-1}$ &  2.25 10$^{-4}$\\
 311&  1.60&    98&  6.00 10$^{-1}$ &  9.50 10$^{-1}$ &  4.80 10$^{-1}$ &  1.17 10$^{-3}$\\
 312&  1.88&   115&  4.42           &  5.60           &  3.64           &  4.90 10$^{-3}$\\
 313&  1.95&   119&  6.45           &  8.11           &  5.53           &  6.43 10$^{-3}$\\
\hline
\multicolumn{6}{c}{P=14.49~~~~~ $e$=0.5 ~~~~~~ $r_{per}$=32.1} \\
\hline
 501&  0.20&    13&  5.40 10$^{-1}$ &  2.83           &  3.62           & -1.97 10$^{-3}$\\
 502&  0.40&    26&  2.20 10$^{-1}$ &  1.04           &  3.00 10$^{-2}$ &  2.52 10$^{-3}$\\
 503&  0.60&    39&  1.30 10$^{-1}$ &  4.30 10$^{-1}$ &  6.00 10$^{-2}$ &  9.25 10$^{-4}$\\
 504&  0.80&    52&  4.30 10$^{-2}$ &  9.00 10$^{-2}$ &  2.00 10$^{-2}$ &  1.75 10$^{-4}$\\
 507&  1.00&    66&  2.40 10$^{-2}$ &  1.10 10$^{-2}$ &  4.72 10$^{-3}$ &  1.57 10$^{-5}$\\
 508&  1.05&    69&  2.50 10$^{-2}$ &  8.93 10$^{-3}$ &  5.53 10$^{-3}$ &  8.50 10$^{-6}$\\
 509&  1.10&    72&  2.80 10$^{-2}$ &  1.03 10$^{-2}$ &  6.76 10$^{-3}$ &  8.85 10$^{-6}$\\
 510&  1.20&    78&  3.90 10$^{-2}$ &  3.01 10$^{-2}$ &  8.38 10$^{-3}$ &  5.43 10$^{-5}$\\
 511&  1.40&    92&  1.10 10$^{-1}$ &  2.40 10$^{-1}$ &  3.09 10$^{-2}$ &  5.23 10$^{-4}$\\
 512&  1.60&   105&  4.30 10$^{-1}$ &  1.13           &  1.60 10$^{-1}$ &  2.43 10$^{-3}$\\
 513&  1.88&   123&  4.65           &  8.18           &  3.55           &  1.16 10$^{-2}$\\
 514&  1.95&   128&  8.80           &  13.2           &  7.42           &  1.44 10$^{-2}$\\
\hline
\multicolumn{6}{c}{P=31.17~~~~~ $e$=0.7 ~~~~~ $r_{per}$=32.15} \\
\hline
 701&  0.20&    14&  3.60 10$^{-1}$ &  3.07           &  4.96 10$^{-3}$ &  7.66 10$^{-3}$\\
 702&  0.40&    28&  2.00 10$^{-1}$ &  1.40           &  3.86 10$^{-2}$ &  3.40 10$^{-3}$\\
 703&  0.60&    42&  1.20 10$^{-1}$ &  5.10 10$^{-1}$ &  5.89 10$^{-2}$ &  1.13 10$^{-3}$\\
 704&  0.80&    56&  4.70 10$^{-2}$ &  1.20 10$^{-1}$ &  1.12 10$^{-2}$ &  2.72 10$^{-4}$\\
 707&  1.00&    70&  2.20 10$^{-2}$ &  2.43 10$^{-2}$ &  3.02 10$^{-3}$ &  5.32 10$^{-5}$\\
 708&  1.05&    73&  2.06 10$^{-2}$ &  2.09 10$^{-2}$ &  2.87 10$^{-3}$ &  4.51 10$^{-5}$\\
 709&  1.10&    77&  2.05 10$^{-2}$ &  2.40 10$^{-2}$ &  2.69 10$^{-3}$ &  5.33 10$^{-5}$\\
 711&  1.40&    98&  6.90 10$^{-2}$ &  4.20 10$^{-1}$ &  1.11 10$^{-2}$ &  1.02 10$^{-3}$\\
 712&  1.60&   112&  3.00 10$^{-1}$ &  2.21           &  8.33 10$^{-2}$ &  5.32 10$^{-3}$\\
 713&  1.88&   131&  3.71           &  16.2           &  2.24           &  3.49 10$^{-2}$\\
 714&  1.95&   136&  5.66           &  22.6           &  3.70           &  4.72 10$^{-2}$\\
\hline
\hline
\end{tabular}
\tablefoot{Surface equatorial rotation velocity V$_{rot}$ is in km/s; total energy dissipation rate 
averaged over the orbital cycle, total energy dissipation rate at periastron and at apastron are in units 
of 10$^{35}$ ergs/s.  The last column gives ($\dot{E}_{tot}^{per}$-$\dot{E}_{tot}^{ap}$)/L$_*$ where 
L$_*$=4.$\times$10$^{37}$ ergs/s is the adopted total luminosity for a 10$_\odot$ star such as the one 
we model.  A negative value occurs when the brightness at apastron is greater than at periastron.  The 
heading for each set models lists the corresponding orbital period in days, eccentricity, and orbital separation at periastron $R_{per}$ in solar units.
}
\end{table*}

\begin{table}[h]
\centering
\caption{Circular orbit set. \label{table_circular}}
\begin{tabular}{ l r l l l }
\hline\hline
Model&  $\beta_0$ & V$_{rot}$&$<\dot{E}_{tot}>$& $\Delta \dot{E}$/L$_*$    \\
\hline
\multicolumn{5}{c}{P=6\,d, e=0}   \\
\hline
 102  &0.40  &18  &0.24                   &6.0$\times$10$^{-4}$  \\
 103  &0.60  &27  &4.2$\times$10$^{-2}$  &1.0$\times$10$^{-4}$ \\
 104  &0.80  &36  &5.8$\times$10$^{-3}$  &1.4$\times$10$^{-5}$ \\
 106  &0.95  &43  & 7.9$\times$10$^{-5}$  &2.0$\times$10$^{-7}$ \\
 108  &1.05  &48  & 6.8 $\times$10$^{-5}$ &1.7$\times$10$^{-7}$ \\
 109  &1.20  &55  & 4.3 $\times$10$^{-3}$ &1.1$\times$10$^{-5}$ \\
 111  &1.60  &73  & 0.17                  &4.3$\times$10$^{-4}$ \\
 113  &1.95  &89  & 0.92                  &2.3$\times$10$^{-3}$ \\
\hline
\hline
\end{tabular}
\tablefoot{Surface equatorial rotation velocity V$_{rot}$ is given in km/s;
total energy dissipation rate averaged over the orbital cycle in units of 10$^{35}$
ergs/s.}
\end{table}

\subsection{Input parameters}

The TIDES input parameters are described in Table~\ref{table_modelsmi}, where we also list the
values that were kept constant. The masses and the primary star's radius
were chosen to be similar to those used in \citet{2009ApJ...704..813H}, \citet{2016A&A...590A..54H},
and \citet{2021A&A...653A.127K}.  The eccentricities ($e$=0, 0.1, 0.3, 0.5, 0.7) were chosen to sample
the range observed in heartbeat stars reported by Kolaczek-Szym\'anski et al. (2021).  The
orbital periods were chosen so as to keep the periastron distance at an approximately constant value,
$r_{per}$=32.1 R$_\odot$.  The synchronicity parameters  $\beta_0$=0.2 - 1.95 were chosen to sample 
the range in values of the eclipsing binaries studied
by  \citet{2017AJ....154..250L} (their Fig. 7)\footnote{It is important to note that the rotation velocities
quoted in \citet{2017AJ....154..250L} were determined based on the assumption that the light variations are
caused by starspots.}  

Table~\ref{table_overview} provides an overview of the parameters that were held constant for each
eccentricity set: Orbital period P, orbital semi-major axis $a$, number of orbital periods 
and phases that were sampled, N$_{cy}$ and N$_{phases}$, respectively. 
Detailed descriptions for each model are provided in Table~\ref{table_eccentric} 
for the eccentric orbit models and Table~\ref{table_circular} for the circular orbit models.
Column 1 lists the model number, column 2 the value of $\beta_0$ and
column 3 the surface equatorial rotation velocity.  This velocity corresponds to the initial equatorial rotation
velocity of the uniformly rotating star. 





\section{Results \label{section_results}}

An initial uniform rotation state transitions into one in which the average rotation rate of each 
layer is either faster (if $\beta_0<1$) or slower (if $\beta_0>1$) than the initial rate
\citep{2021A&A...653A.127K}.  In all cases, the surface layer changes its velocity
faster than any of the other layers. As a  result, the average angular rotation is a function of
distance from the stellar center.  Furthermore, because the equatorial latitude is the one most strongly 
affected by the tidal force, it slows down or speeds up (depending on the $\beta_0$ value) faster than 
the polar regions.  Thus, there is also an average differential rotation structure in the polar direction.  
Both of these differential rotation structures  contribute to the total energy dissipation rate when a 
turbulent viscosity is present.  In addition, the models rapidly develop the tidal bulges with
the corresponding inflow and outflow of material. This relative motion of different mass parcels 
also contribute to the total energy dissipation rate.

\subsection{Pseudosynchronicity}

\begin{figure}
\centering
\includegraphics[width=0.48\columnwidth]{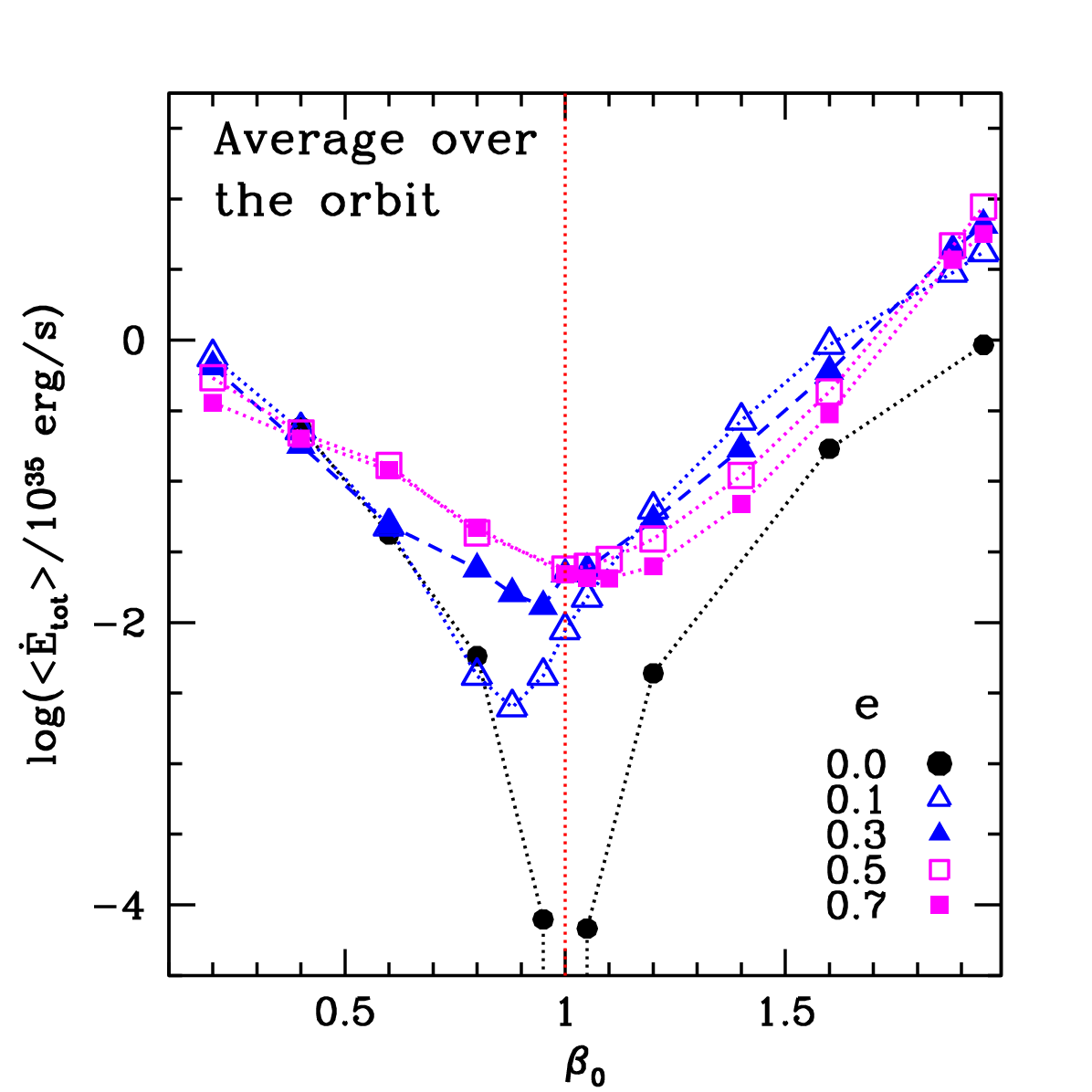}
\includegraphics[width=0.48\columnwidth]{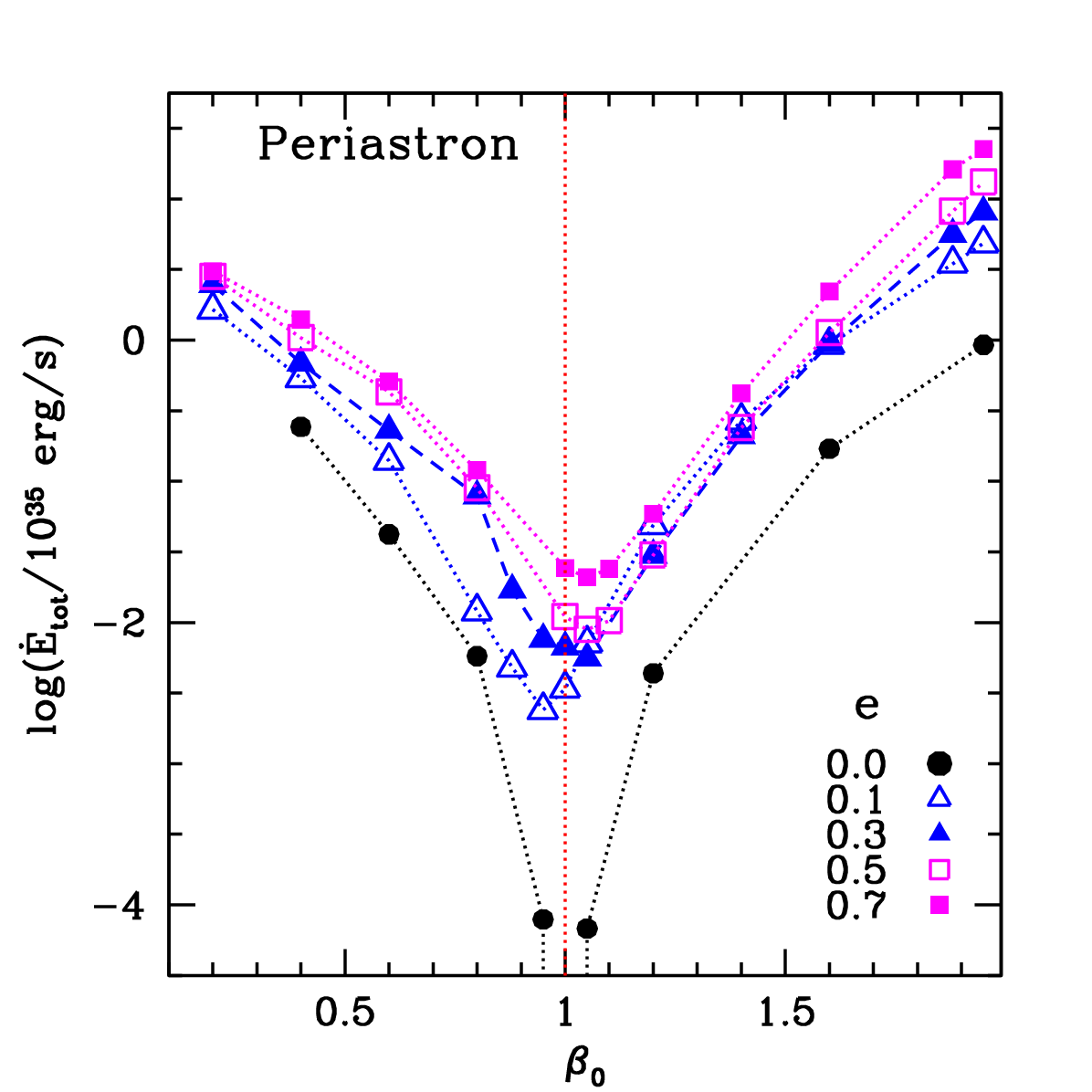}
\caption{Energy dissipation rate as a function of the degree of synchronicity $\beta_0$. Colors correspond 
to the eccentricity of the model, as indicated in the legend. {\it Left:} Average over an orbital cycle,
$\left<\dot{E}_{tot} \right>$. {\it Right:} Value at periastron, $\dot{E}_{tot}^{per}$.  The vertical
line indicates $\beta_0$=1, corresponding to the synchronous rotation rate for a circular orbit system.
\label{prompot_plots} }
\end{figure}

The energy dissipation rate averaged over an orbital period $\left<\dot{E}_{tot}\right>$ is
listed in Table~\ref{table_eccentric} (column 3), as are the instantaneous energy dissipation rates 
$\dot{E}_{tot}$ at periastron (column 4) and apastron (column 5). This table clearly shows 
that larger departures from synchronicity lead to larger $\left<\dot{E}_{tot}\right>$ values. As
shown in Fig.~\ref{prompot_plots}, minimum dissipation rates occur near the synchronization rotation 
rate $\beta_0 \simeq$1  but,  except for $e$=0, the actual minimum does not occur at $\beta_0$=1
but at the pseudosynchronous rate, $\beta_{ps}$, as shown analytically by \citet{1981A&A....99..126H}. 
However, contrary to the analytical result which predicts a relatively constant $\beta_{ps}$
value for $e\geq$0.2, our $\beta_{ps}$ values depend on eccentricity.  Another difference is that although
the analytical result predicts a unique $\beta_{ps}<$1 value for the minimum energy rotation state, 
our  $e$=0.5 and 0.7 models show a  flat and extended minimum at $\beta_0>$1.  
A similar result was found by \citet{Moreno:2011jq} based on  denser $\beta_0$ grids. The flat minimum 
is also consistent with the recent finding of a blurred pseudosynchronization in numerical 
calculations of the high eccentricy binary KOI-54 \citep{2023ApJ...953...48T}.

A bifurcation between subsynchronous and supersynchronous rotation is present in the perturbed rotation
velocity.  This is quantified by finding the maximum/minimum angular velocity at each latitude and
layer depth, and is  defined as $\pm |\omega_{max}|$.  That is, the value of the maximum departure 
from the rest frame velocity, with the sign indicating whether it is in the direction of rotation 
(positive sign) or in the opposite direction. In the subsynchronous rotators, $|\omega_{max}|>$0, as 
expected since the action of the tidal force is to accelerate the stellar rotation toward corotation.  
The opposite is true for the supersynchronous rotators.  Examples of this behavior for the surface layer
at the equator are shown in Figs. \ref{fig_omdp_e0.1and0.3} and \ref{fig_omdp_e0.5}.  These show that
when rotation is close to the pseudosynchronous rate, $|\omega_{max}|$ displays high frequency oscillations 
in addition to the rapid excursions to either large positive or negative values around periastron.
For systems with circular orbits, the energy dissipation rate decreases symmetrically around $\beta_0$=1,
tending toward a null value at synchronicity as expected (Table~\ref{table_circular}).

\subsection{Orbital phase-dependence and periastron events}

\begin{figure}
\centering
\includegraphics[width=0.48\columnwidth]{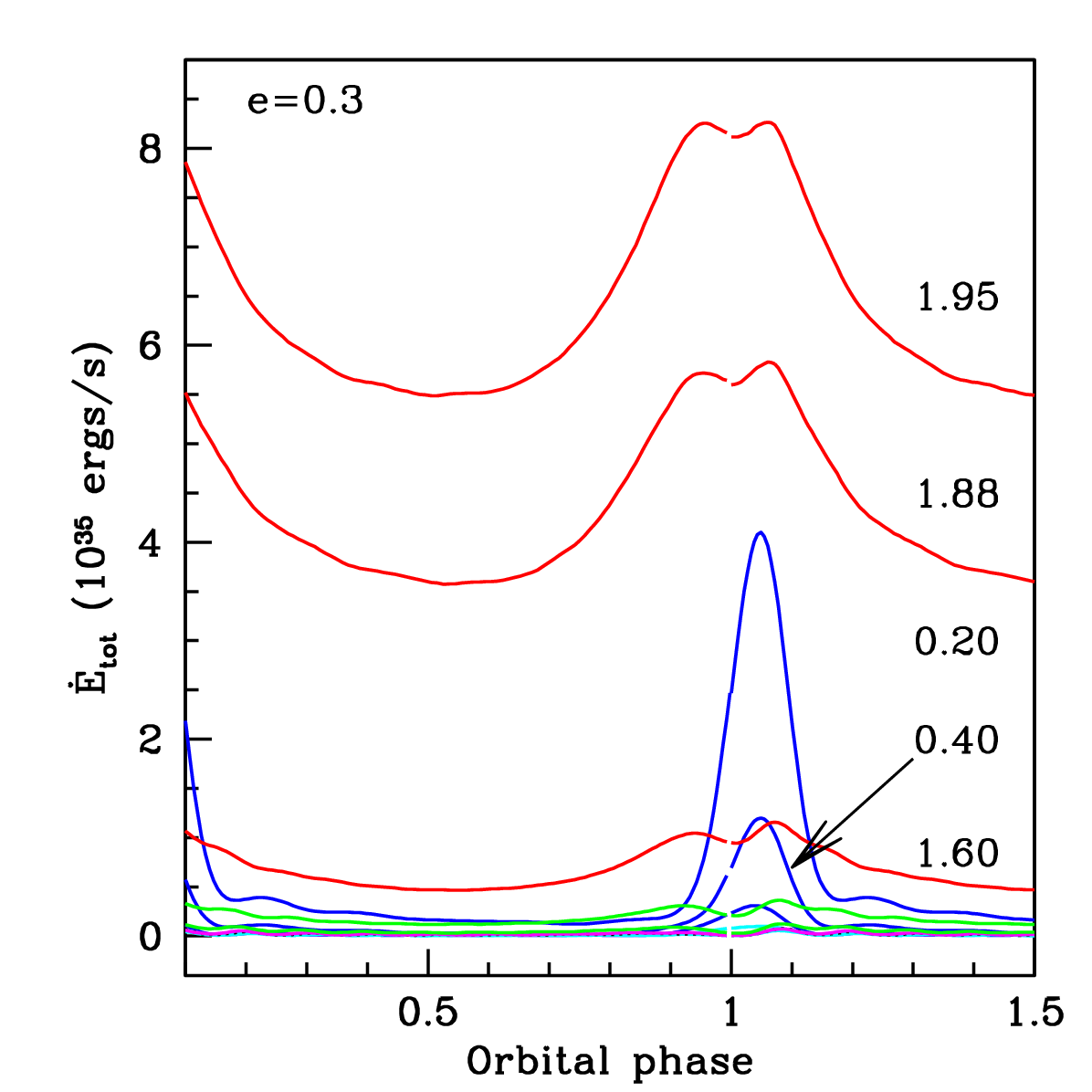}
\includegraphics[width=0.48\columnwidth]{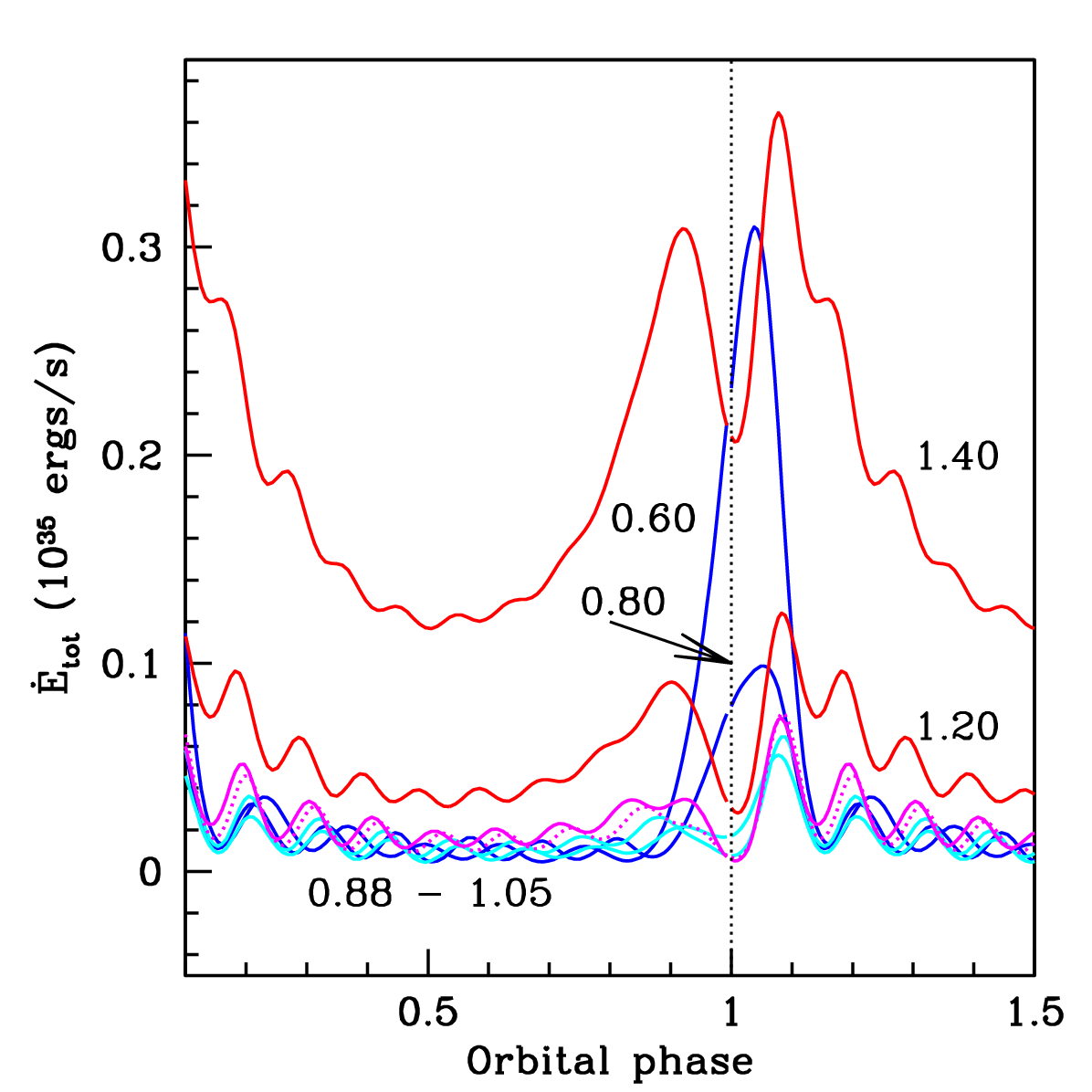}
\caption{Total energy dissipation rate (in units of 10$^{35}$ ergs/s) as a function of the orbital 
phase for the $e$=0.3 set of models.
Models with $\beta_0<$0.80 (blue) show a relatively narrow periastron effect with a single peak centered
in phase slightly after periastron passage. Models with $\beta_0\geq$1.20 show a significantly
broader periastron effect with a structured peak. Models with 1.20$\leq\beta_0>$0.80        
tend to have a dominant peak after periastron and multiple weaker peaks at other phases.
{\it Left:} All models of the set; {\it Right:} Models with 0.60$\geq\beta_0\leq$1.40, which have 
the weakest energy dissipation rates and are plotted on a scale that shows the 
variations associated with the tidally excited oscillations.}
\label{fig_heartbeat2}
\end{figure}

\begin{figure}
\centering
\includegraphics[width=0.48\columnwidth]{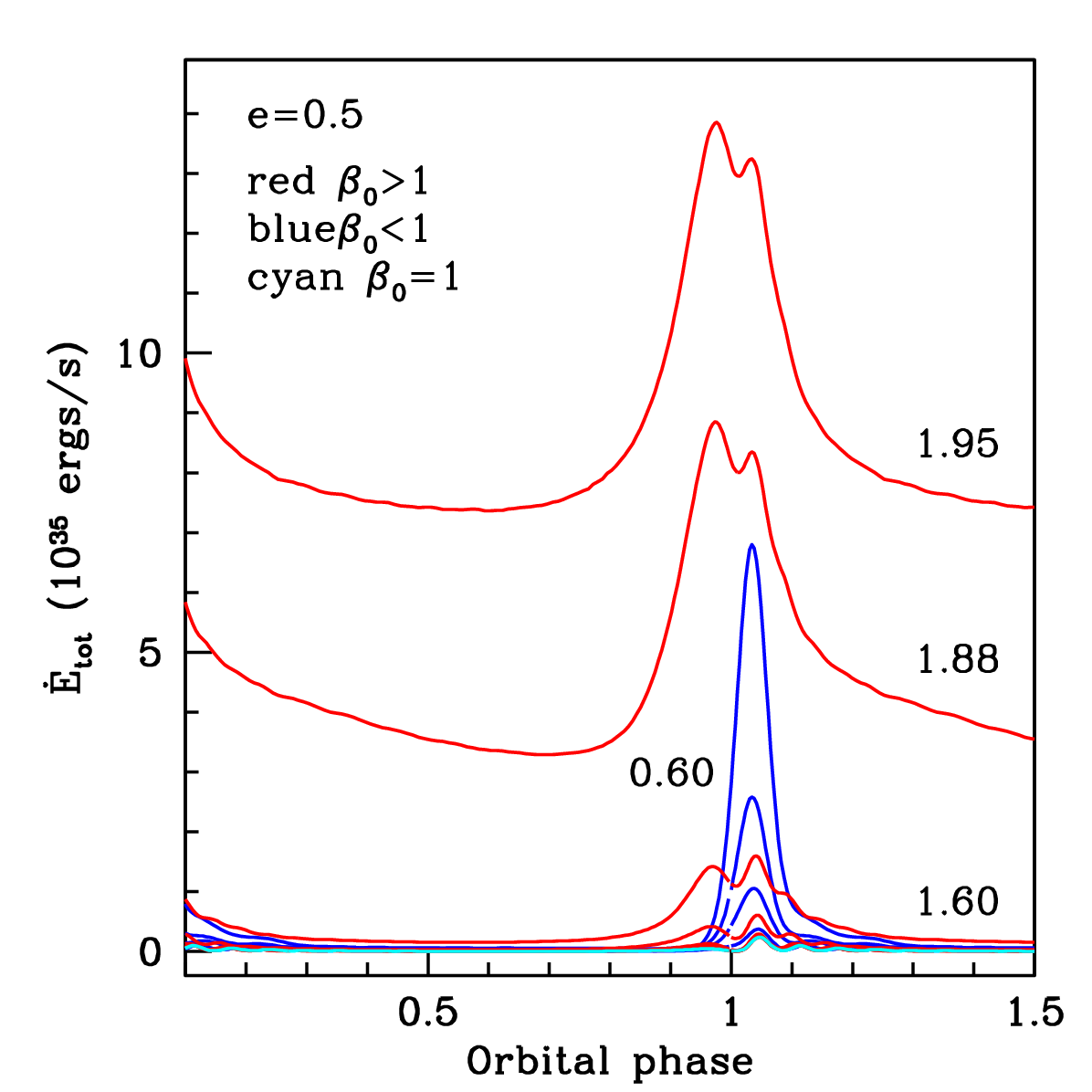}
\includegraphics[width=0.48\columnwidth]{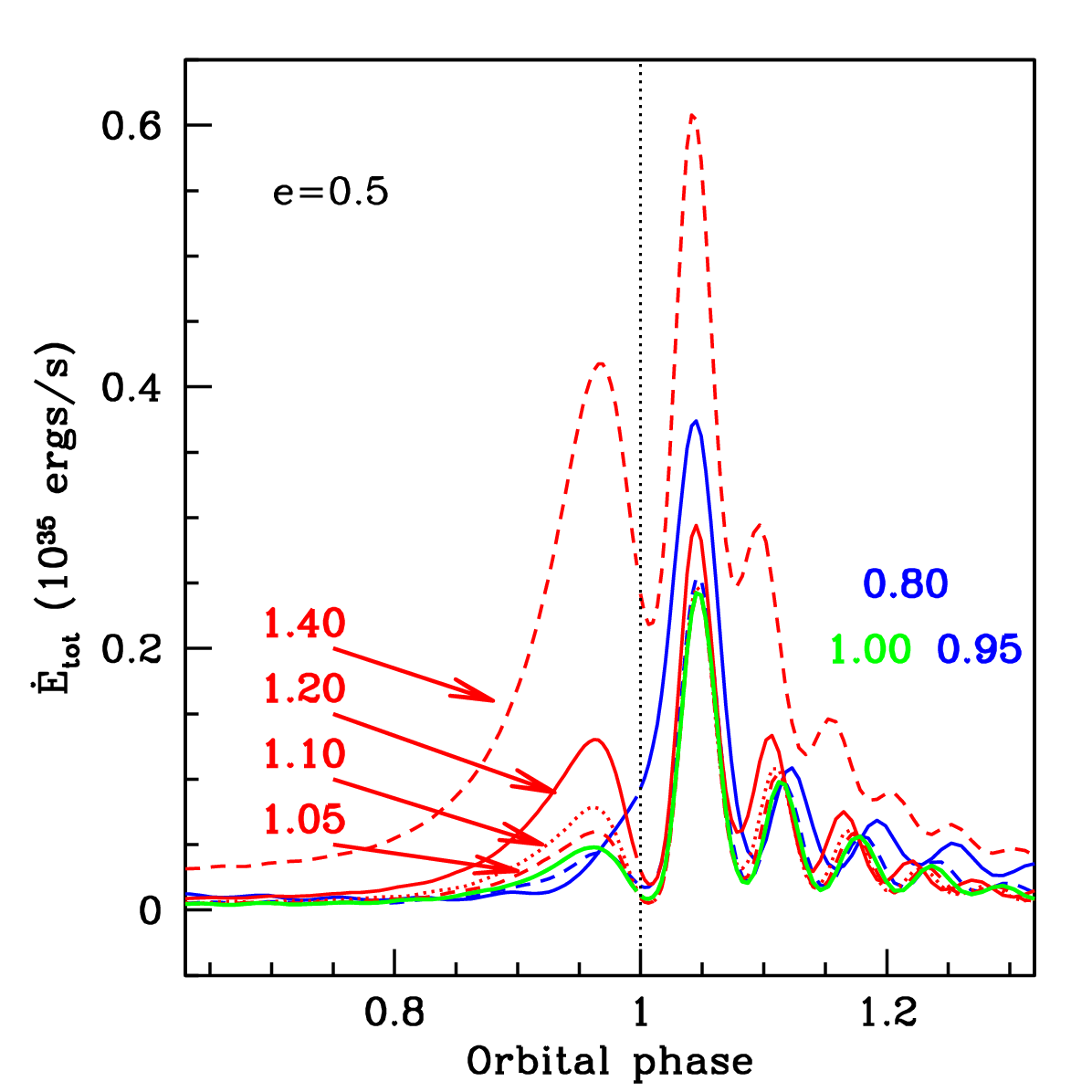}
\includegraphics[width=0.48\columnwidth]{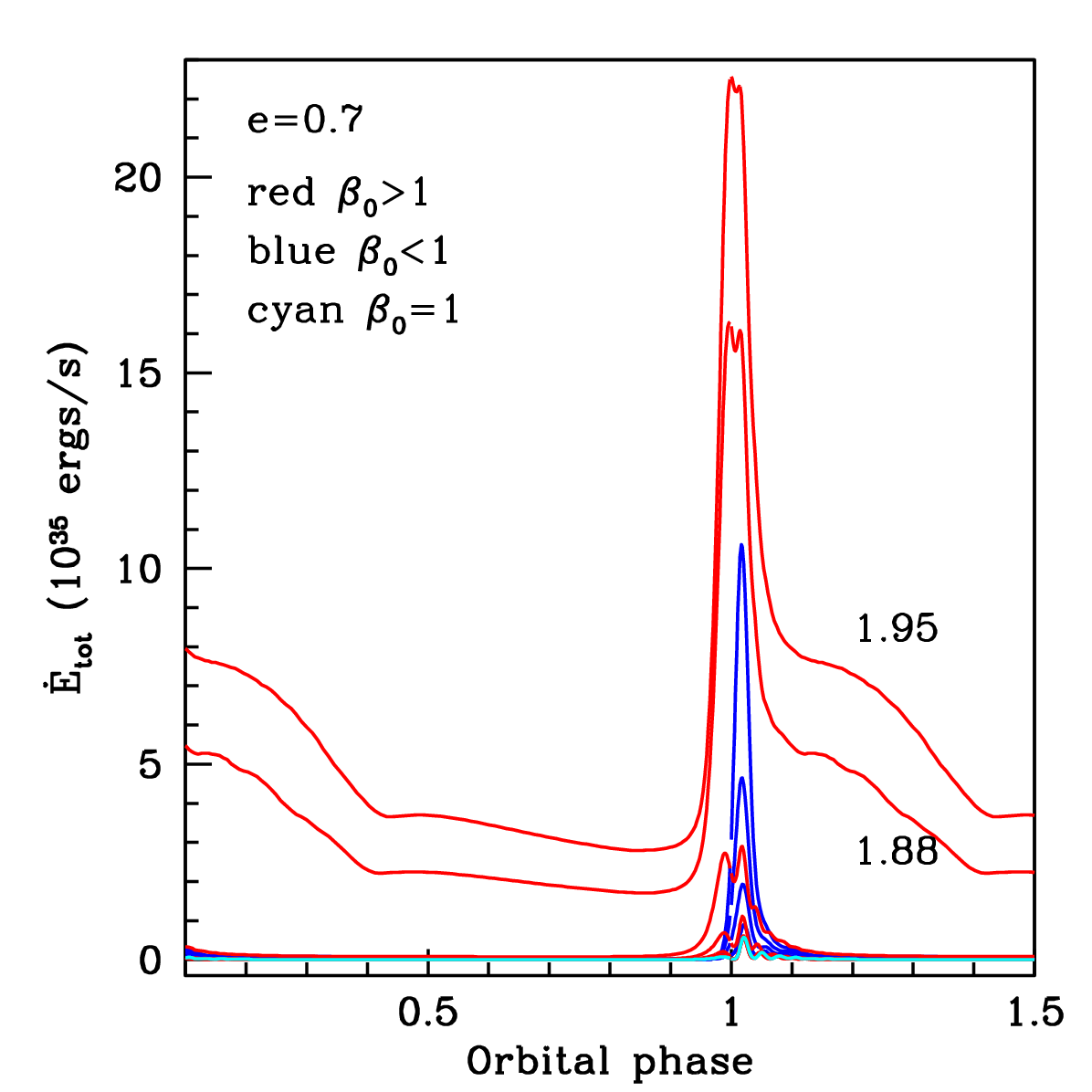}
\includegraphics[width=0.48\columnwidth]{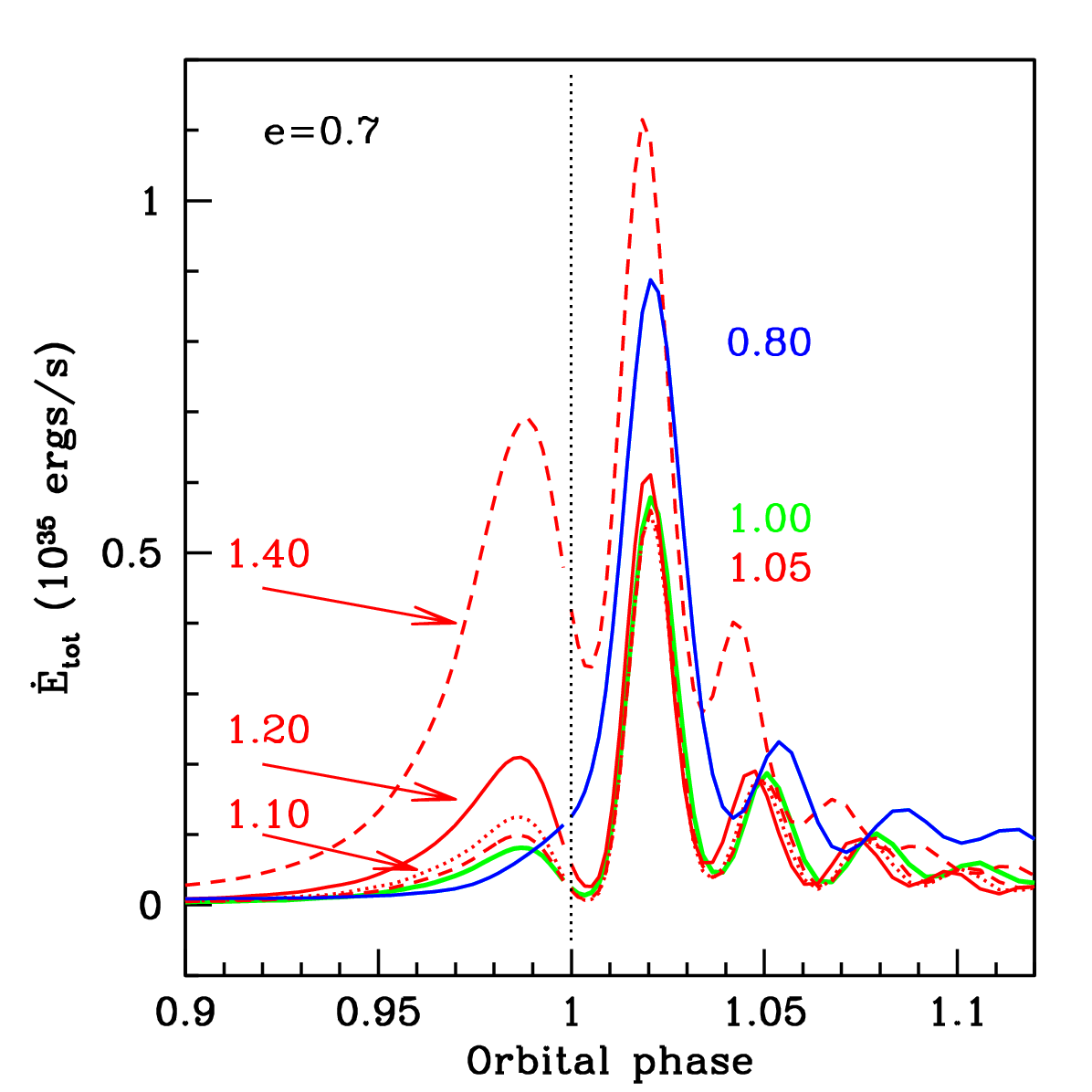}
\caption{Total energy dissipation rate (in units of 10$^{35}$ ergs/s) as a function of the orbital 
phase for the $e$=0.5 (top) and $e$=0.7 (bottom) models.  All $\beta_0$ values modeled are 
plotted on the left while on the right we zoom in on the 0.80$\leq\beta_0\leq$1.40 cases. 
Although the orbital elements are all the same in these models, the perturbation duration 
around periastron differs, being longer for faster rotation rates (larger $\beta_0$ values). 
The lowest-lying curves are those for $\beta_0$=1.00 (green), 1.05 (red dash) and 1.10 (red dot), 
all of which display several discrete peaks between orbital phase 0.9-1.3.
}
\label{fig_heartbeat3}
\end{figure}

Fig.~\ref{fig_heartbeat2} illustrates the orbital phase-dependent behavior of $\dot{E}_{tot}$
in the $e$=0.3 models.  Several trends are clearly visible: a) there is always an increase in 
$\dot{E}_{tot}$ around the time of periastron; b) maximum intensity tends to occur shortly after 
periastron; c) the durations of the periastron event is shorter in the subsynchronous models than 
in the supersynchronous models; d) models with $|\beta_0-1| \simeq$1 present a dip at
periastron;  and e) models with $|\beta_0-1| \simeq$1 display high frequency oscillations.
The orbital phase-dependent behavior for $e$=0.5 and $e$=0.7 is similar to that of $e$=0.3,
as shown in Fig.~\ref{fig_heartbeat3}, but the amplitude of the rise in brightness around
periastron increases with increasing eccentricity.   

Assuming that the primary star in our models is similar to the B-type star in the {\it Spica} 
system, we can adopt $L_*$=4. 10$^{37}$ ergs/s \citep{2016MNRAS.458.1964T} and  compare
the brightness enhancements at periastron due to tidal shear energy dissipation to this luminosity.
We list in Column 6 of Table~\ref{table_eccentric} the difference between the maximum around 
periastron and the mean value at apastron normalized to $L_*$:
$\Delta \dot{E}_{tot}$/L$_*$=($\dot{E}_{tot}^{per}-\dot{E}_{tot}^{ap}$)/L$_*$.  
We find values mostly in the range 10$^{-5}$ to 10$^{-3}$, with a few as low as 10$^{-6}$ and 
others as high as 10$^{-2}$, the latter in models with the highest rotation rates and largest eccentricity.
Values in the range 10$^{-6}$-10$^{-3}$ are typically found in the heartbeat stars.

A curious feature in Column 6 of Table~\ref{table_eccentric} is that some of the values of 
$\Delta \dot{E}_{tot}$/L$_*$ are negative.  This is because of the dip at periastron which, in models
with $|\beta_0-1| \simeq$1 can have a lower energy around apastron than the value of the mean.

The value of viscosity $\nu_{turb}$ plays a crucial role in determining the energy dissipation
rate. Our results confirm the idea that $\nu_{turb}$ has a time-dependence in addition to a dependence on 
spatial coordinates.  For example, we find $\nu$ in the range 10$^{12}$ cm$^2$/s - 10$^{15}$ cm$^2$/s  
for the inner and surface layers, respectively (see Table~\ref{table_viscosities}) at periastron and 
smaller values around apastron.  

The turbulent viscosities we derive for inner layers are on the order of those found by
\citet{2007ApJ...655.1166P} and others \citep{2008EAS....29...67Z} 10$^{10}$ - 10$^{13}$ cm$^2$/s,
but the surface values are likely to be overestimates, an issue that could be resolved with a 
variable $\lambda$ value if a straightforward theoretical formulation existed for predicting the 
rate of conversion from the tidal kinetic energy into heat.  This remains an open question.


\subsection{Conclusions from this section}

We find that the pseudosynchronous rotation rate $\beta_{ps}$ for $e$=0.1 and 0.3 is slower  
than the synchronous rate at periastron while that for $e$=0.5 and 0.8 is faster.  Contrary to
the results of \citet{1981A&A....99..126H}, $\beta_{ps}$ is not approximately constant 
for eccentricities $e>$0.2 (see his Fig. 3, the curve labeled $\Omega_{ps}$/$n_p$, which is
equivalent to our $\beta_{ps}$).  A possible source for this discrepancy may be the assumption of
uniform rotation in \citet{1981A&A....99..126H} while our results are obtained for a star that
has differential rotation.  Our results are consistent with those of \citet{Moreno:2011jq} who 
found that  $\beta_{ps}$ does not have a unique value for $e$=0.5 and 0.7 but that the minimum 
energy state occurs for a range of rotation rates near the synchronous rate at periastron.  This 
broad $\beta_{ps}$ region is also  consistent with the blurred pseudosynchronization found in the  
numerical calculations of \citet{2023ApJ...953...48T}.

There are clear trends in the orbital phase-dependent behavior of the energy dissipation
rates visible in all eccentricities that we modeled:  a) maximum brightness tends to occur slightly 
after periastron, not at periastron;  b) the duration over which an increased brightness may be 
observed is shorter in the subsynchronous models than in the supersynchronous models; c) for fixed
values of orbital period and eccentricity,  the energy dissipation rate  increases with increasing 
departure from synchronicity, with the largest values found for $\beta_0>>$1;  d) for fixed values
of $\beta_0$ and periastron distance,  the energy dissipation rate  increases with increasing
eccentricity; e) models with $|\beta_0-1| \simeq$1 present a dip in brightness at periastron 
as well as prominent high frequency variations that we identify with tidally excited oscillations.


%

\begin{table}[h!]
\caption{Input parameters and results MACHO 80.7443.1718. \label{table_MACHO}}
\centering
\begin{tabular}{ l c l }
\hline\hline
Parameter&Value&Notes\\
\hline
\hline
\multicolumn{3}{c}{Data from observations and input parameters}              \\
\hline
P$_{orb}$                  & 32.83627\,d           &  (a)                               \\
e                          & 0.507                    &  (a)                      \\
m$_{1}$                    &34.5 M$_{\odot}$ &  (b)                        \\
m$_{2}$                    &15.7 M$_{\odot}$ &  (b)                      \\
R$_{1}$                    &23.7 R$_{\odot}$ &  (b)                     \\
V$_{rot}$                  &250 km/s         &   (c)               \\
L$_*$                      &1.57$\times$10$^{39}$ ergs/  &  (b)                      \\
$\Delta$F$_{peri}$/F       &$\sim$23\%                &  (d)   \\
$\beta_0$                  &1.95                      &        \\                                        
$\lambda$                  & 1                        &         \\                       
N$_{cy}$                   & 100                      &            \\
N$_{r}$                    &5                         &      \\
$a$                        & 159 R$_\odot$            &         \\
$r_{per}$                  & 78.5 R$_\odot$            &         \\
\hline
\multicolumn{3}{c}{Output values and results}               \\
\hline
$\dot{E}_{peri}$           & 3.7$\times$10$^{38}$ ergs/s &  (e)          \\
$\dot{E}_{apo}$            & 6.6$\times$10$^{36}$ ergs/s &  (e)          \\
$\left<\dot{E}_{tot}\right>$ &  6.5$\times$10$^{37}$  ergs/s   & (e)           \\
$\Delta \dot{E}_{peri}$/L$_*$  & 23.5\%           &     (f)        \\
$R_{max-peri}$                &23.83 R$_{\odot}$ &   (g)  \\
$R_{min-per}$                 &23.63 R$_{\odot}$ &   (g)  \\ 
$R_{max-ap}$                  &23.76 R$_{\odot}$ &   (g)  \\
$R_{min-ap}$                  &23.66 R$_{\odot}$ &   (g)  \\ 
$\left<R_{per}\right>$        &23.71 R$_{\odot}$ &   (h)\\
$\left<R_{ap}\right>$         &23.70 R$_{\odot}$ &   (h)\\
\hline
\hline
\end{tabular}
\tablefoot{(a) \citet{2019MNRAS.489.4705J}; (b) \citet{2021MNRAS.506.4083J}; (c) Adopting
an orbital inclination $i$=44$^\circ$ and $\vv \sin(i)$=174 km/s from \citet{2021MNRAS.506.4083J};
(d) estimated from Fig. 9 of \citet{2021MNRAS.506.4083J} which shows maximum light at levels of
$\sim$1.2 to $\sim$1.26 (depending on the TESS sector) above the average value $\sim$1 between 
subsequent periastron passages. ; (e) Output from TIDES-nvv; 
(f) ($\dot{E}_{tot}^{peri}$-$\dot{E}_{tot}^{apo}$)/L$_*$;
(g) Maximum and minimum equatorial radius at periastron and apastron, including high frequency
oscillations, from the TIDES-1 version of the code with $\nu$=0.028 R$^2_\odot$/d.  The
corresponding values from the TIDES1-vv model with $\lambda$=0.01 are  23.76 and 23.67 R$_\odot$.
(h) average polar radius at periastron and apastron from the TIDES-1 version of the code.}
\end{table}

\section{Modeling the heartbeat star MACHO 80.7443.1718}

MACHO 80.7443.1718 is located in the Large Magellanic Cloud Lucke-Hodge 58 association, which
contains stars formed within the past few million years \citep{1994AJ....108.1256G}. It was 
classified as an eclipsing binary system \citep{1997ApJ...486..697A} and has recently been found 
to display a $\sim$40\% peak-to-peak variability amplitude with morphology that places it among 
the heartbeat-type stars \citep{2019MNRAS.489.4705J}. An analysis of the light curve and
radial velocity curve of photospheric absorption lines  was performed by \citet{2021MNRAS.506.4083J},
leading to the conclusion that minimum light coincides with the time of periastron passage, defined 
to occur at orbital phase $\varphi$=0, and maximum light at $\varphi\sim$0.04.    
The times of minima in the light curve were found to undergo a systematic change by 11.1 s/yr 
\citep{2022A&A...659A..47K}. If interpreted to represent  times of central eclipse, this 
implies an extraordinarily large rate of change in the orbital period.

The observed brightness increase $\Delta$F/F$\sim$23\% at periastron far exceeds the typical values in the
range 10$^{-6}$ - 10$^{-3}$ usually found in heartbeat stars \citep{2017MNRAS.472.1538F}.
\citet{2023NatAs...7.1218M} pointed out that the stellar radius of 24 R$_\odot$  deduced 
from the spectroscopic observations \citep{2021MNRAS.506.4083J} was inconsistent with the significantly
larger tidal distortion needed by standard theories to explain such strong light curve variations.
In order to address this problem, \citet{2023NatAs...7.1218M} performed a hydrodynamic simulation 
that predicts a tidal wave raised at periastron having an extension of several solar radii, and 
concluded  that the breaking wave could lead to shock-heating on the surface of the star.  These
results, however, are based on several assumptions, the strongest of which is that the core of the
primary star is excised from the calculation and treated as a point mass leaving the envelope
detached from the core.  In addition, both the radius and orbital eccentricity are overestimated
to obtain the desired result.  

In the previous sections, we showed that tidal shear energy dissipation can play an important 
role in describing the heartbeat-star phenomenon.  In this section we present
results from applying the model to MACHO 80.7443.1718.

\subsection{Brightness variations from tidal heating}

We computed a TIDES-nvv model for MACHO 80.7443.1718 using as input the values that were obtained 
by \citet{2021MNRAS.506.4083J} from their analysis of the observational data. Table~\ref{table_MACHO}
lists these input parameters.  The projected rotation velocity deduced from the widths of 
the photospheric absorption lines is 174 km/s which, with an assumed orbital inclination of 44$^\circ$,
implies an equatorial rotation velocity V$_{rot}$=250 km/s. This yields $\beta_0$=1.95. The range 
in V$_{rot}$ values implied by the measurement uncertainties in V$\sin$(i) quoted by 
\citet{2021MNRAS.506.4083J} corresponds to a range $\beta_0$=1.57-2.34.  Hence, the primary star of 
the system is in supersynchronous rotation for all possible V$_{rot}$ values.  We adopt $\beta_0$=1.95 
for the calculation, which means that at apastron it is even more highly supersynchronous, 
$\beta_{ap}$=18.2.  We used the same computational parameters as in Table~\ref{table_modelsmi} except 
for the number of layers that we set to $N_r$=5 since the largest tidal shear energy dissipation 
rates occur in the surface layers.  The model was evolved out to 100 orbital periods (9 years), 
with output retained at  5, 10, 20 and 50 orbital periods.  

\begin{figure}
\centering
\includegraphics[width=0.98\columnwidth]{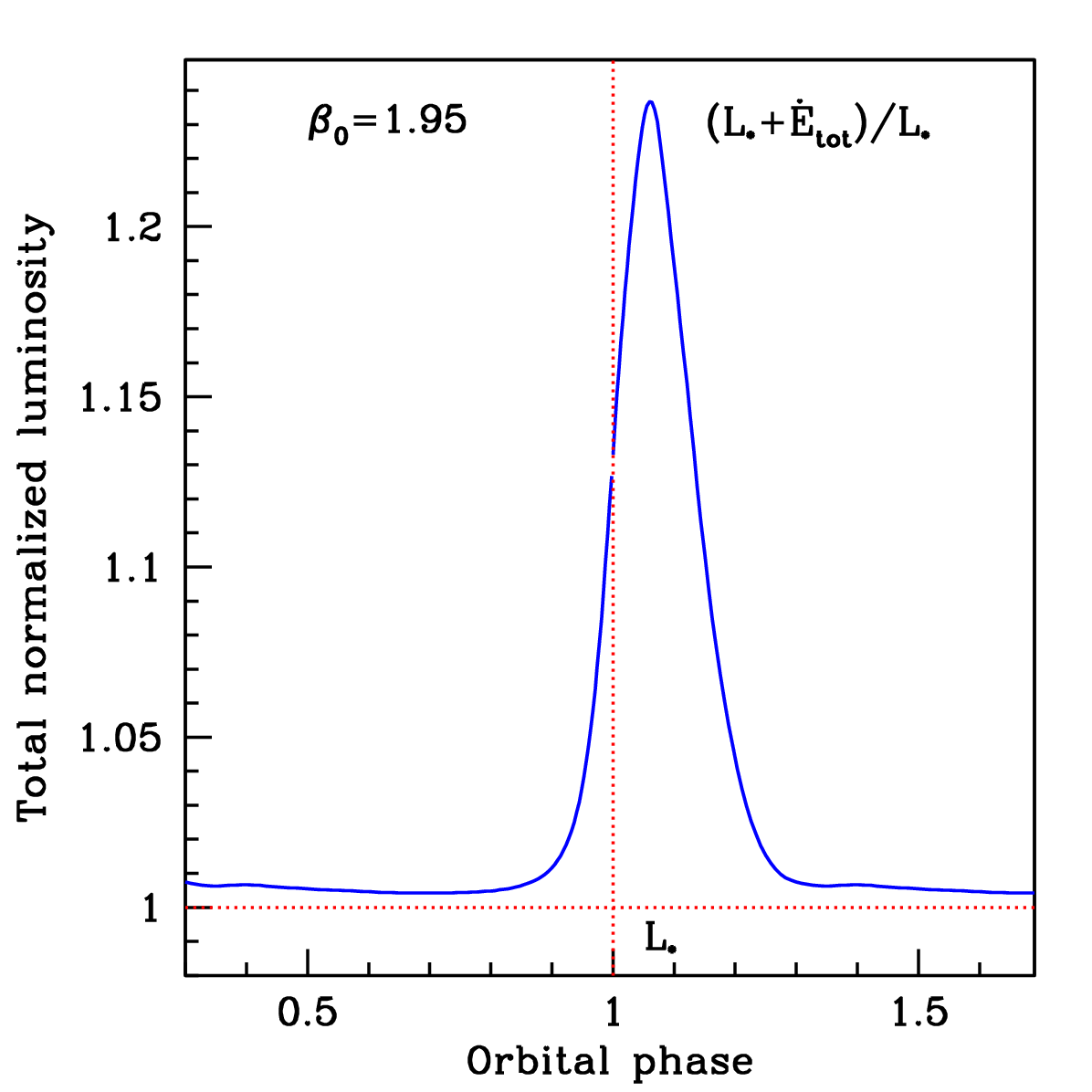}
\caption{Predicted MACHO 80.7443.1718 light curve from the tidal shear energy dissipation calculation
added to the stellar luminosity given by J21, L$_*$=1.63 10$^{39}$ ergs/s, and normalized to this
value. The vertical dotted line indicates the time of periastron and the horizontal line indicates
the normalized luminosity level.  Tidal shear contributes $\sim$23\% of the luminosity peak around the
time of periastron.
\label{pt35_loudest_adding_Lstar} }
\end{figure}

The resulting total tidal energy dissipation rate $\dot{E}_{tot}$ at periastron and apastron is,
respectively, 3.7 10$^{38}$ ergs/s and 6.6 10$^{36}$ ergs/s.  The normalized light curve,
obtained by adding the phase-dependent $\dot{E}_{tot}$ to the stellar luminosity L$_*$=1.57 10$^{39}$ 
ergs/s cited in \citet{2021MNRAS.506.4083J} and dividing by this number, is illustrated in 
Fig.~\ref{pt35_loudest_adding_Lstar}. It shows a 23\% increase at maximum, with maximum occurring 
at orbital phase $\sim$0.06 after periastron passage. The brightness increase is not symmetrical 
around periastron but initiates at $\varphi\sim$0.9 and ends at $\varphi\sim$0.3.  These results are 
all consistent with the observations.

We emphasize that our results follow directly from a calculation performed from first principles, with
no fitting to the data, no prior assumptions and no adjustable parameters other than the conversion 
rate of kinetic energy into heat ($\lambda$). However, several comments are in order.
The n-layer, variable-viscosity version of our model provides only the tidally induced 
perturbations in the azimuthal direction due to computational challenges\footnote{With n-layers, the 
absence of a lower boundary for each layer allows vertical displacements of the volume elements that 
exceed their vertical extension; that is, they can become detached from each other, a condition 
that halts the numerical integration.  A full SPH computation is required to keep track of the 
3D location of each volume element and allow for buoyancy effects.}  Thus, it is likely to yield an 
underestimate of the total energy dissipation rate.   On the other hand, the value of $\lambda$ we
have used is the upper limit of this parameter and is certainly an overestimate.
Adding in the energy dissipation rate associated with the radial perturbation components might 
counterbalance the use of a smaller, more realistic $\lambda$ value.   

A final comment is in order regarding the average tidal energy dissipation rate that we find in
our model which, at first sight, might seem very large in the context of the tidal evolution of
the system.  Using Eq. (9) of \citet{2016RMxAA..52..113K} and the energy dissipation rate average over 
an orbit, we find $\dot{P}/P \simeq$1.4$\times$10$^{-3}$ s/yr.  \citet{2022A&A...659A..47K} reported 
variations in the light curve that they interpret in terms of a decreasing orbital period at a rate of 11 s/yr,
which would imply an energy loss from the orbit that is orders of magnitude larger than that
associated with the tidal shear energy dissipation rate that we have derived.

\begin{figure}
\centering
\includegraphics[width=0.98\columnwidth]{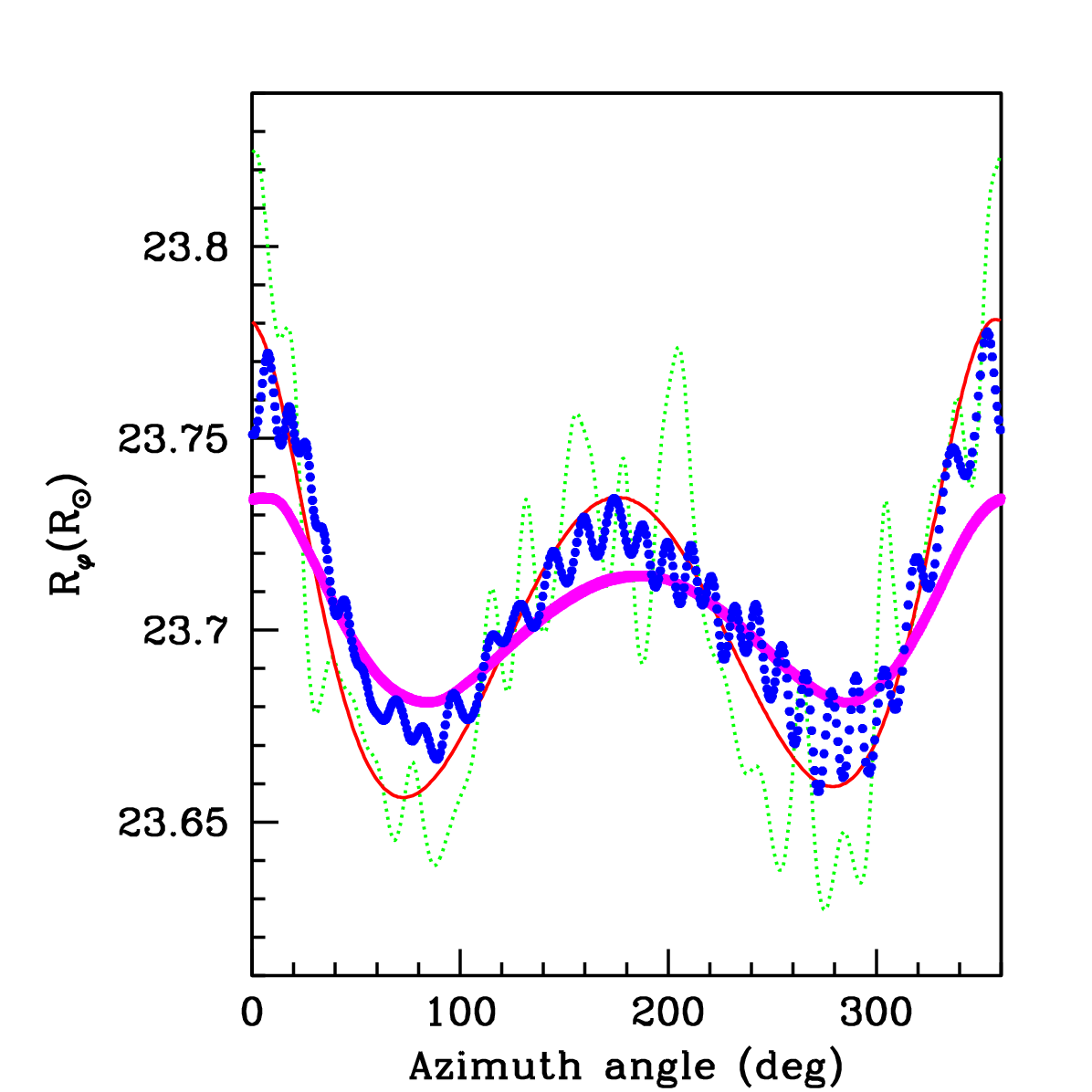}
\caption{Equatorial radius of MACHO 80.7443.1718 at periastron as a function of the longitude obtained with
the TIDES-1 models using $\nu$=0.028 R$^2_\odot$/d (green) and $\nu$=2 R$^2_\odot$/d (red), and
the TIDES-1vv with $\lambda$=0.01 (blue) and $\lambda$=1 (magenta). The sub-binary longitude
is at $\varphi$=0$^\circ$. Each dot in the plot correspond to a grid point. 
The smaller viscosity models display 28 wave-like oscillations in the radius over the
2$\pi$ radians of the stellar circumference.
\label{propaz_rphi_loudest} }
\end{figure}

\begin{figure}
\centering
\includegraphics[width=0.48\columnwidth]{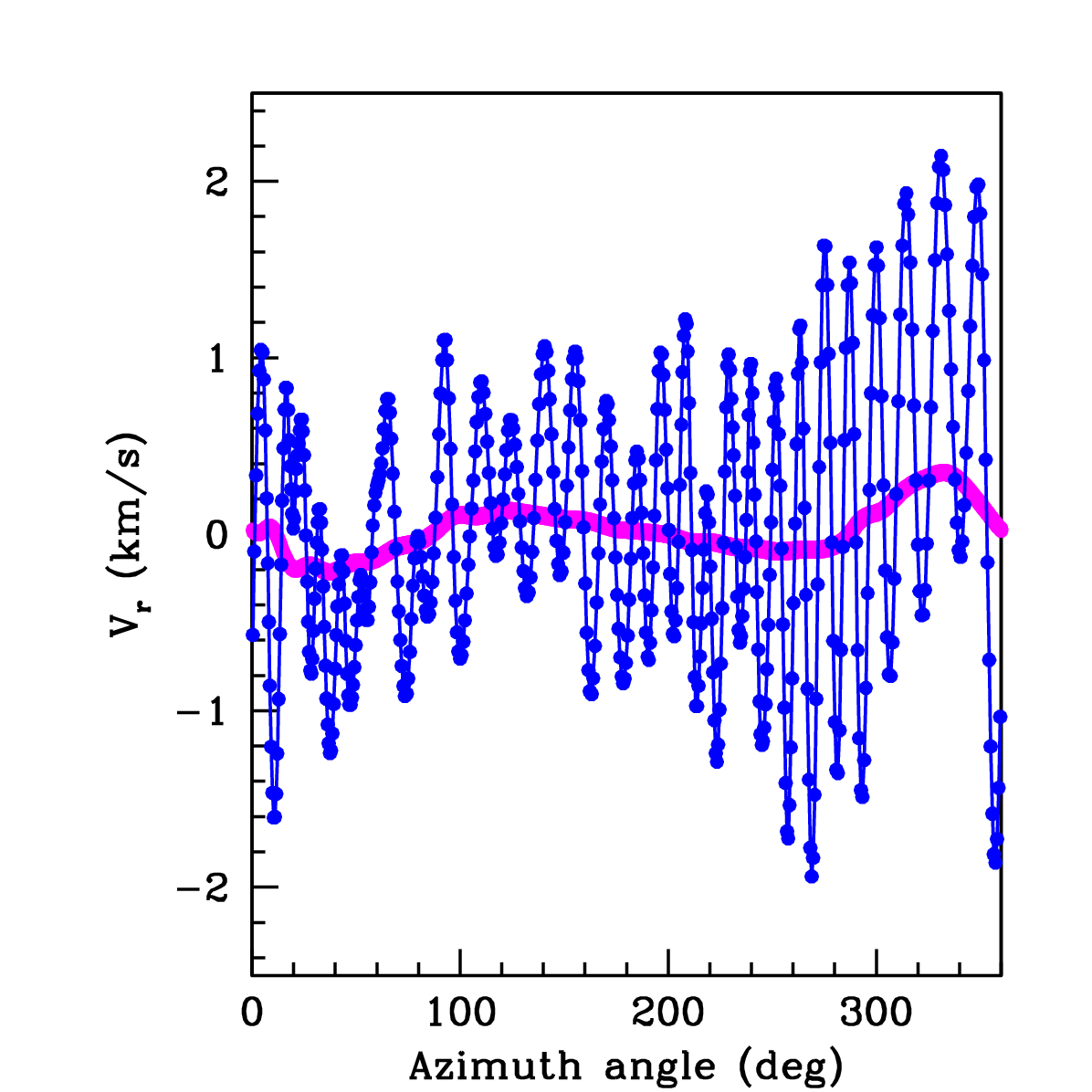}
\includegraphics[width=0.48\columnwidth]{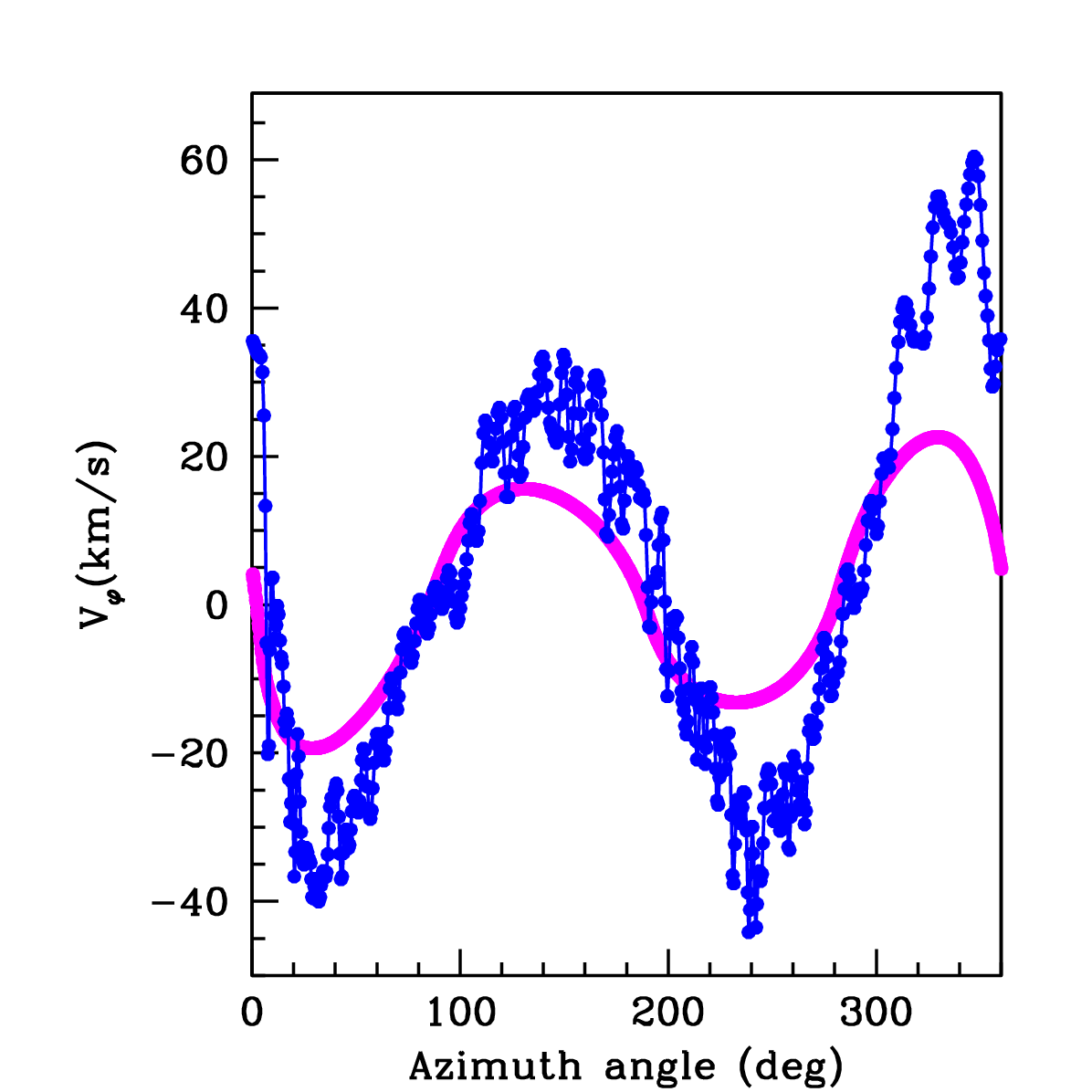}
\caption{Velocity in the radial direction (left) and azimuthal direction (right) along
the equator of the MACHO 80.7443.1718 surface at periastron as a function of the longitude.
The colors indicate results from the calculations using the code TIDES-1vv with $\lambda$=0.01 (blue); 
and with $\lambda$=1 (magenta). The sub-binary longitude
is at $\varphi$=0$^\circ$. Each dot in the plot correspond to a grid point. The smaller
viscosity models ($\lambda$=0.1) display wave-like velocity structures along the stellar
circumference.
\label{propaz_vphi_loudest} }
\end{figure}

\subsection{Radius and perturbed radial velocity component}

In order to explore the effects of the radial component of the tidal perturbation   
we performed two tests with a one-layer model based on the TIDES-nvv algorithm but that allows
radial motions.  As with the original version, in this new version (TIDES-1vv) the surface layer 
remains attached to the underlying rotating core at all times and is coupled to it
through the viscous force.  The first test was performed with  $\lambda$=1, as was used with the
n-layer calculation and the second  with $\lambda$=0.01. For comparison, we also computed two 
models using the original one-layer model discussed in \citet{Moreno:2005cq} and \citet{Moreno:2011jq}.  
All input parameters are the same as for the TIDES-1vv calculation except that in this case a fixed 
viscosity value is needed as input.  We explored values $\nu$=0.028 R$_\odot^2$/d (1.6 10$^{15}$ cm$^2$/s) 
and $\nu$=2 R$_\odot^2$/d (2.3 10$^{17}$ cm$^2$/s) which are on the order of magnitude of those that
were computed by TIDES-nvv with $\lambda$=1. All the calculations were performed with 500 partitions 
in the azimuthal direction.

We find that the primary tidal bulge at periastron extends no more than $\sim$0.13 R$_\odot$ above 
the equilibrium radius.  The four calculations yield a similar average shape 
(Fig.~\ref{propaz_rphi_loudest}).  The differences can be traced to the viscosity values, with smaller
viscosities allowing for the appearance of high frequency oscillations in both radius and velocity.

The size of the main tidal bulge found in our calculation can be compared with  
the estimate (see, for example, \citet{2008EAS....29...67Z}:

\begin{equation}
\frac{\delta R}{R_1} \simeq \frac{m_1}{m_2} \left(\frac{R_1}{a}\right)^3,
\end{equation}

\noindent where $R_1$ is the equilibrium radius, $m_1$ and $m_2$  the masses of the primary star and 
the perturber, respectively, $a$ is the semimajor axis, and $\delta R=R_{bulge}-R_1$.  This relation 
gives $\delta R\simeq$0.17 R$_\odot$, which is comparable to our maximum value 0.13 R$_\odot$.

In general, the tidal velocity field has components in the three spatial directions.  Our
one-layer computations calculate the values in the radial and azimuthal directions.  The 
two TIDES-1vv tests yield the behavior of the radial component that is shown in 
Fig.~\ref{propaz_vphi_loudest} (left).  The $\lambda$=0.01 calculation displays high frequency 
oscillations with maximum amplitudes $\sim$2 km/s at longitudes 300$^\circ$ - 360$^\circ$.  
Thus,  our model indicates that the rise to the tidal bulge is through a series of oscillations 
of increasing amplitude.  The azimuthal velocity component  also displays high frequency 
oscillations which, however, have a small amplitude compared to the large-scale equilibrium tide component 
which reaches speeds as high as $\sim$100 km/s as it rises into the primary tidal bulge 
(Fig.~\ref{propaz_vphi_loudest} right).

The most significant difference between the one-layer calculations is the presence of high frequency
oscillations in the models having smaller viscosity values.  This suggests that although large
$\lambda$ values are needed for our results to be consistent with the observed maximum brightness, 
the value of this parameter needs to decrease at the surface in order to allow for TEOs, a reasonable
expectation given the significantly lower density of the surface compared to inner layers, as well
as its optically thinner properties allowing energy to escape.  Allowing for a variable $\lambda$
parameter is, however, beyond the scope of the current paper.\footnote{We note that the percentage
contribution from each layer of radius $r$/R$_\odot$=[16.6,18.0,19.4,20.9,22.3] to the total energy 
dissipation rate at periastron is respectively [$<$0.01, 0.2, 0.55, 0.2, 0.05].  Hence, a very
small $\lambda$ value in the surface layer will not affect the conclusion that tidal shear
energy dissipation can explain the brightness variations of MACHO 80.7443.1718 and other heartbeat stars.}

\begin{figure}
\centering
\includegraphics[width=0.98\columnwidth]{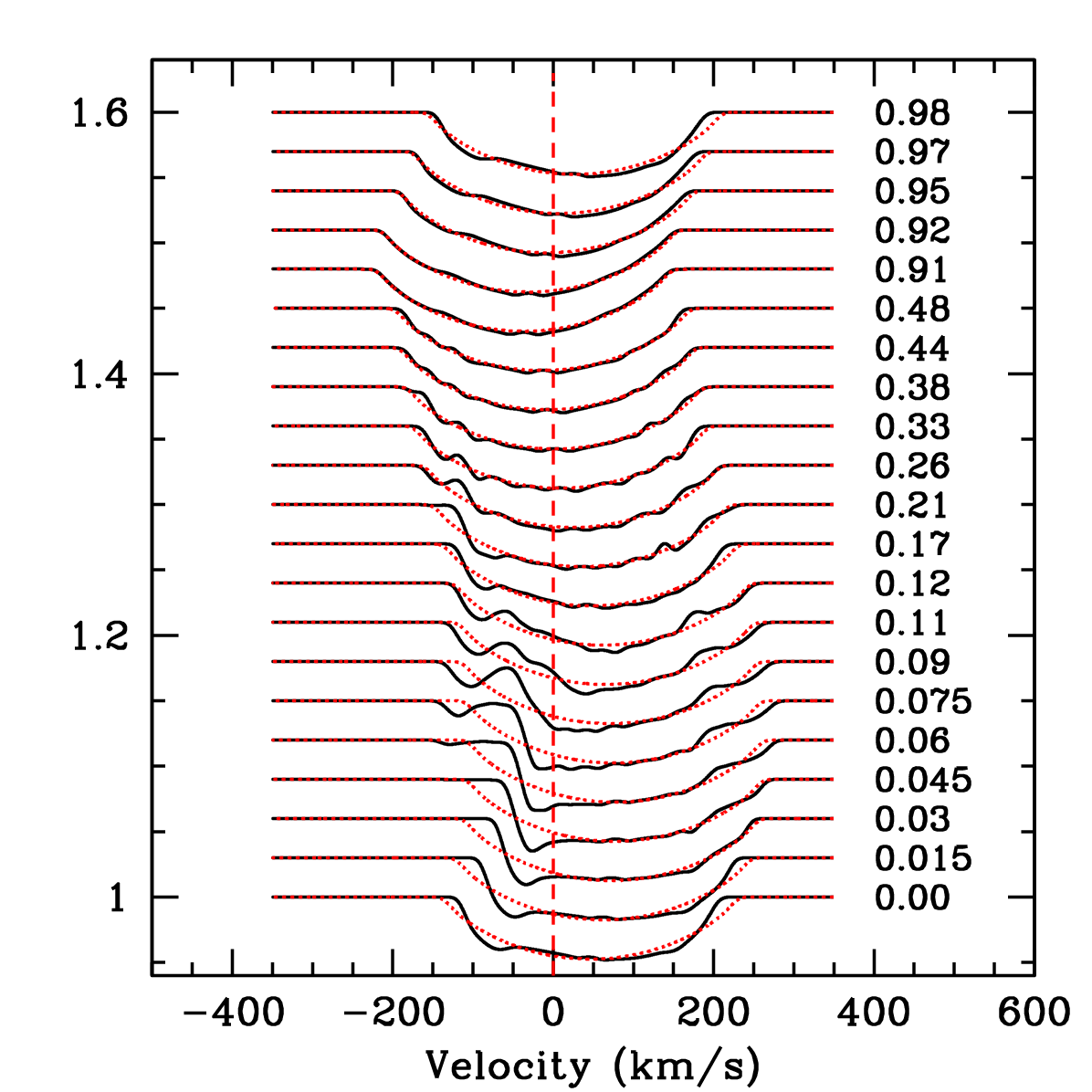}
\caption{Line profiles over an orbital cycle computed with the one-layer model of Moreno et al. (2005)
for  MACHO 80.7443.1718 assuming $\beta_0$=1.95 and the masses, radius, and orbital elements given by
Jayasinghe et al. (2021).  The unperturbed and rotationally broadened profiles (red dots) are superposed on the
predicted profiles (black) which show significant distortions only in the orbital phase between just before
periastron ($\varphi\sim$0.97) until $\sim$0.3 in phase after periastron, consistent with the duration
of the luminosity increase.
\label{line_profiles}}
\end{figure}

\subsection{Photospheric line profile variations}

The tidal flows and radial motions excited by the tidal perturbation alter the smoothly varying
rotation velocity across the stellar disk, projected onto the line of sight to the observer.  
Hence, instead of the typical rotationally broadened photospheric absorption line profile that 
would be observed if the surface were stationary in the rest
frame, it appears distorted and varies over time as different portions of the stellar surface come
into view.  \citet{Moreno:2005cq} incorporated the calculation of line profiles into the TIDES-1
calculation.  The tidally perturbed velocities on the visible portion of the stellar disk are projected 
onto the line of sight to the observer.  The corresponding Doppler shifts are applied to the local
(on each surface element) line profile and all line profiles are added to produce  the photospheric line 
profile that is predicted for each particular orbital phase.  The method has successfully modeled the
line profile variations that are observed in the eccentric B-type binary {\it Spica} 
\citep{2009ApJ...704..813H, 2013A&A...556A..49P, 2016A&A...590A..54H}.

Fig.~\ref{line_profiles} is a montage of line profiles predicted by the TIDES-1 calculation using
the input parameters for MACHO 80.7443.1718 in Table~\ref{table_MACHO}.  They are stacked from bottom to 
top in order of increasing orbital phase and show that the strongest perturbations are expected to be observed 
just before periastron and until approximately 0.3 in phase after periastron.  It would be interesting to
determine whether similar variations are indeed present in lines that are not affected by a stellar wind, 
such as \ion{He}{1} $\lambda$ 4921 shown in \citet{2021MNRAS.506.4083J} but, unfortunately, the high
quality spectra presented by these authors only cover the orbital phase range 0.27-0.81.


\begin{figure}
\centering
\includegraphics[width=0.48\columnwidth]{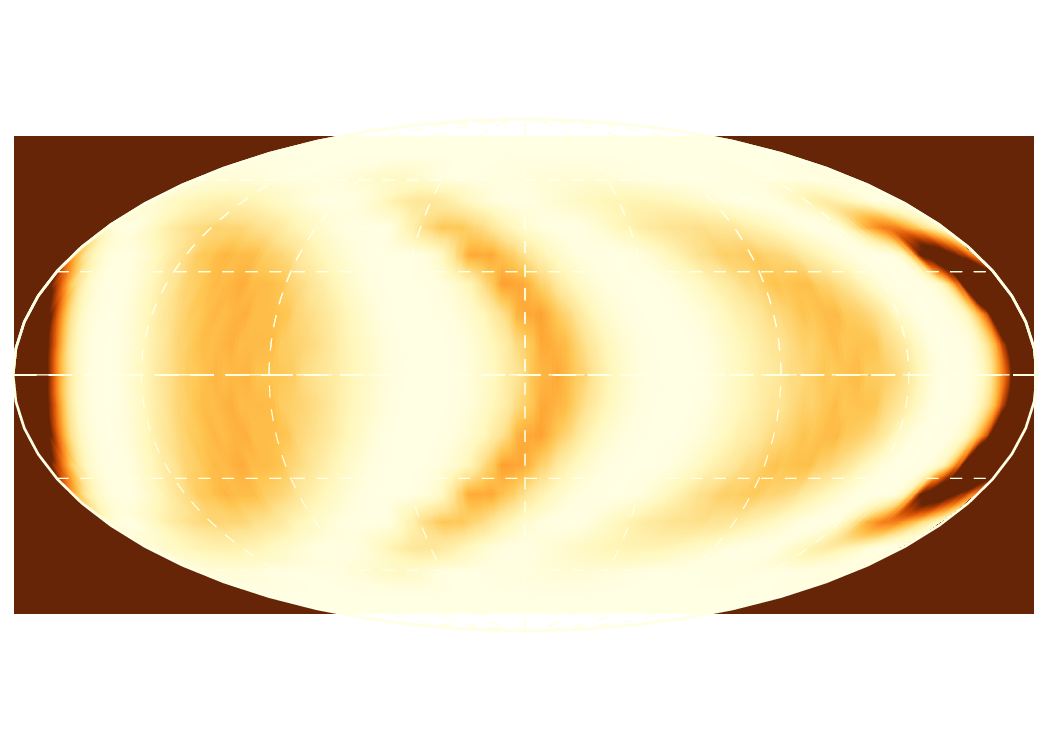}
\includegraphics[width=0.48\columnwidth]{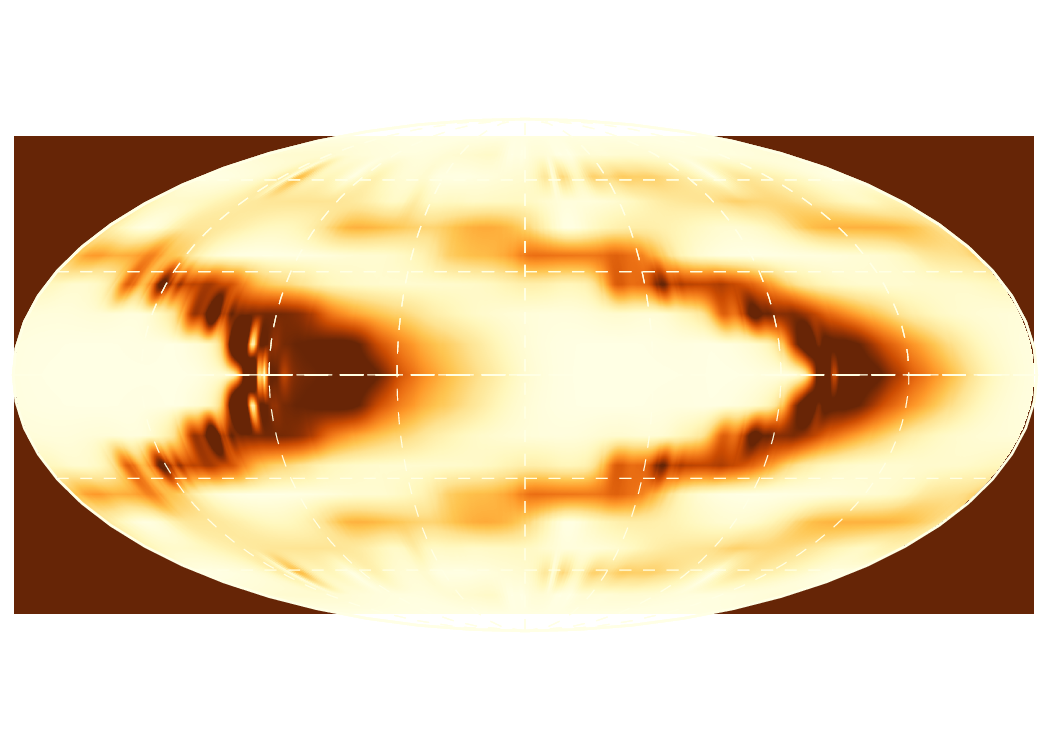}
\includegraphics[width=0.48\columnwidth]{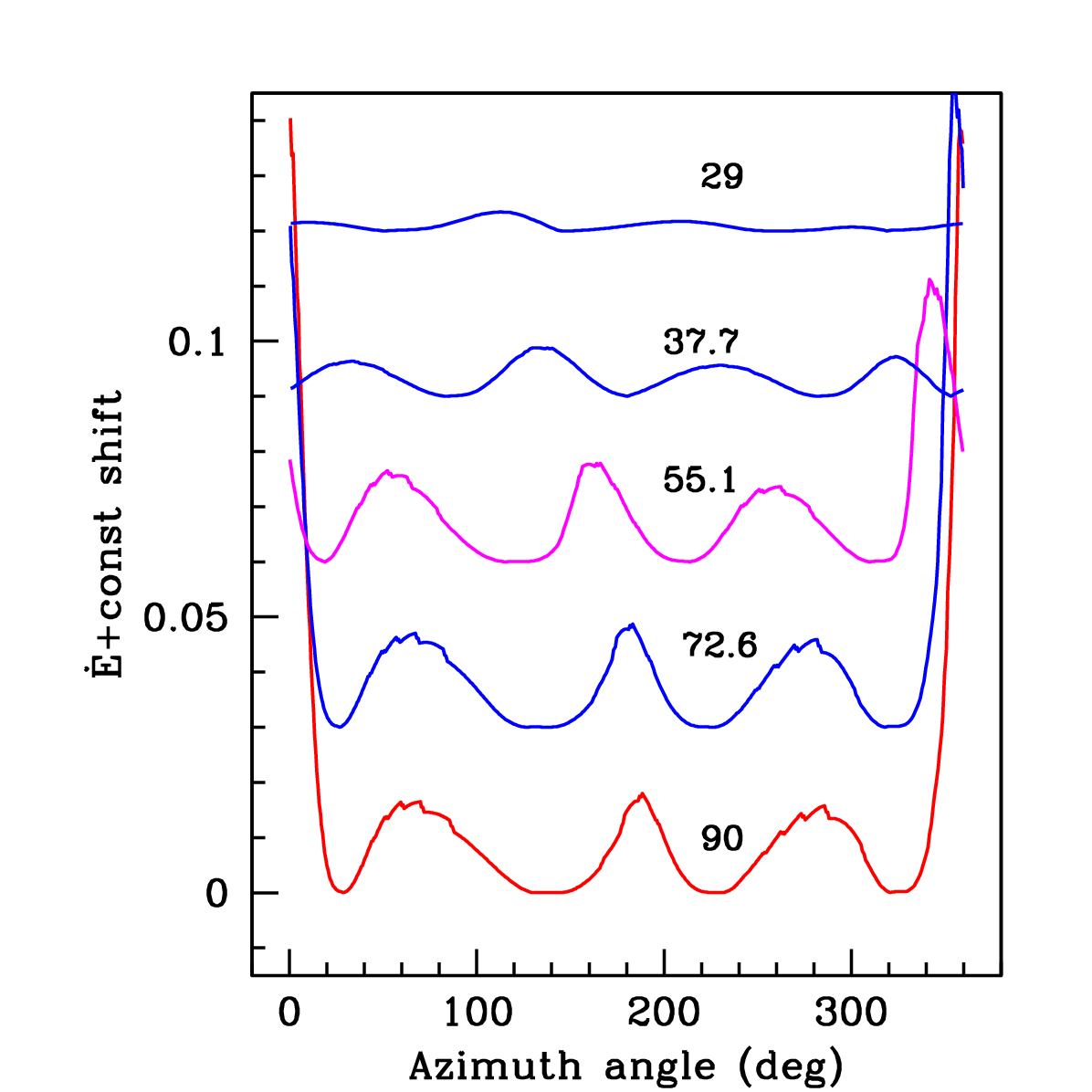}
\includegraphics[width=0.48\columnwidth]{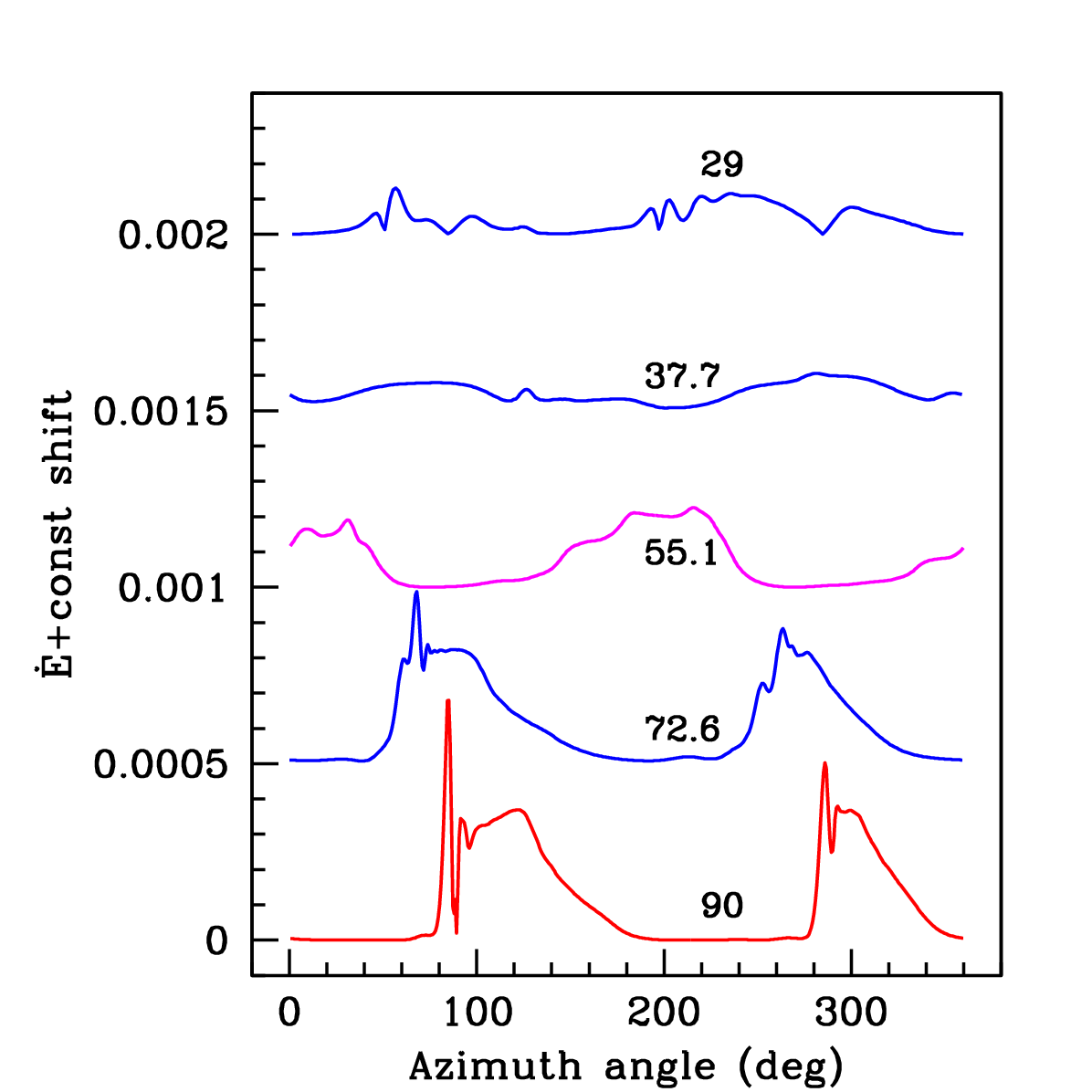}
\caption{Distribution of the tidal shear energy dissipation in MACHO 80.7443.1718 from the 
TIDES-nvv model at periastron (left) and apastron (right).  {Top:} Maps of the surface, with the
sub-binary longitude at $\varphi$=0$^\circ$ and 360$^\circ$ (the rims of the maps),and the poles 
at the top and bottom.  Dark brown represents maximum energy dissipation rate and  minimum is color 
coded white.  {\it Bottom:} Plots of the corresponding $\dot{E}$ per unit volume (in units
of 10$^{35}$ ergs/s/unit volume)  at various latitudes between the equator and the pole.  Each curve is 
labeled with its corresponding colatitude angle. Each latitude is shifted vertically by 0.03 units for 
display purposes.
\label{maps_dotE} }
\end{figure}

\subsection{Non-isotropic energy dissipation structure}

The distribution of the tidal shear energy dissipation rate over the surface is non-isotropic,
and changes between periastron and apastron as shown in Fig.~\ref{maps_dotE}.  At periastron,
maximum energy is concentrated around the sub-binary longitude and extends to colatitudes as
high as  $\sim$60$^\circ$.  Three additional locations of large energy dissipation rates can be
seen at approximately 90$^\circ$, 180$^\circ$ and 270$^\circ$  from the sub-binary longitude.  
At apastron, maxima occur near 90$^\circ$ and 270$^\circ$ at the equator, however at higher 
latitudes maxima are spread out in long arch-like structures spanning broad ranges in longitude.

\subsection{Conclusions of this section}

Our n-layer calculation of the tidal shear energy dissipation rate yields a light curve with
a $\sim$23\% maximum which occurs slightly after periastron passage, as is observed.  These results  follow 
directly from a calculation performed from first principles with no fitting to the data and 
no adjustable parameters other than the conversion rate of kinetic energy into heat ($\lambda$).
The only basic assumption in our treatment is that relative motions excited by the tidal force
gives rise to a turbulent viscosity.

We find that the maximum radial deformation of the star at periastron is $\leq$0.2 R$_\odot$,
contrary to the several R$_\odot$ that was found by \citet{2023NatAs...7.1218M}, and that  
the radial velocity of the material rising into the primary bulge is subsonic, $\sim$2 km/s,
contrary to the 20 km/s found by these authors.  Our results indicate that it is the horizontal 
velocity component that reaches supersonic velocities, attaining speeds as high as $\sim$50 km/s.

The light curve does not predict the minimum that is observed at periastron.  This minimum could
result from a physical eclipse as suggested by \citet{2023NatAs...7.1218M}, but according to the
parameters derived by \citet{2021MNRAS.506.4083J} such an eclipse would be at best a grazing eclipse.
The tidal shear energy dissipation models discussed in Section~\ref{section_results} do predict a minimum
at periastron but only in stars that are rotating close to synchronicity at this orbital phase.
MACHO 80.7443.1718 is rotating at approximately twice the synchronous rate. An alternative scenario for a 
light curve minimum is one in which a puff of dense material is ejected at periastron.  This material
would dim the star until it has expanded enough to become optically thin,  at which time it would become the
source of emission lines such as H$\alpha$ and H$\beta$.  \citet{2023NatAs...7.1218M}  also suggested that 
the activity at periastron could lead to material leaving the star.  

\citet{2021MNRAS.506.4083J} detected spectral variations in MACHO 80.7443.1718 that suggest that, indeed, 
enhanced mass-loss episodes occur. Hence, the prominent dip in the light curve, not predicted by our 
tidal shear energy dissipation model, could be caused by the ejected material.
We speculate that the mechanism for mass ejection may be found in the large velocity gradients 
around periastron combined with the rapid rise in the stellar bulge.  Furthermore, similar proceses
are likely also present in the companion, whose contribution to the observed phenomenon should in 
the future be assessed.

\section{Summary and conclusions}

We examine the role of the often neglected process of tidal shear energy dissipation $\dot{E}$
in shaping the observational properties of stars in which periastron brightening events occur. Our
model consists of a direct calculation from first principles of the motions induced on a
rotating star by a companion in an eccentric orbit. The value of $\dot{E}$ is computed
using a non-isotropic and time-dependent viscosity. We explore the impact of stellar
rotation on $\dot{E}$ in a test binary consisting of a 10 M$_\odot$ primary that is perturbed by a 7 M$_\odot$
companion  with orbital periods in the range 6\,d-31\,d and eccentricities $e$ in the range 0$-$0.7.  

The model grid includes rotation rates at periastron  that are subsynchronous ($\beta_0<$1), 
synchronous ($\beta_0$=1), and supersynchronous ($\beta_0>$1), where  $\beta_0$=$\omega_0/\Omega_0$.
Here, $\omega_0$ is the rotation angular velocity of the inner, rigidly rotating region 
that includes the stellar core and $\Omega_0$ is the orbital angular velocity at periastron.

We find that the shear energy dissipation rate increases 
by factors $\sim$10$^{-6}$ -10$^{-3}$  at periastron, consistent with the majority of observed 
heartbeat stars, and we find the following trends in $\dot{E}$:
a) maximum $\dot{E}$ occurs at orbital phases slightly after periastron, not at periastron;  
b) the timescale over which $\dot{E}$ increases and then decreases is shorter in the 
subsynchronous models than in the supersynchronous models; c) for fixed values of orbital period 
and eccentricity,  $\dot{E}$ grows with increasing departure from $\beta_0 \sim$1,
with the largest values appearing for $\beta_0>>$1;  d) for fixed values
of $\beta_0$ and periastron distance,  models with larger eccentricity have a stronger 
periastron effect; e) models with $|\beta_0-1| \simeq$1 have the lowest dissipation rates, present a 
dip at periastron, and display the relatively prominent high frequency variations that may be identified
with tidally excited oscillations.

Pseudosyncronism is defined as the  minimum energy rotation state in an eccentric binary.  In this
state, $\beta_{ps}$=$\omega_{ps}/\Omega_{ave}$, where $\omega_{ps}$ is the rotation angular velocity  and
$\Omega_{ave}$ is the average orbital angular velocity.  We confirm the classical result that
0.8$<\beta_{ps}<$1.0 for the cases with e$<$0.3. However, in the $e$=0.5 and 0.7 cases, we find 
1.0$<\beta_{ps}<$1.15, contrary to the  results of \citet{1981A&A....99..126H} who 
found an approximately constant $\beta_{ps}\simeq$0.82 for eccentricities $e>$0.2.

In order to compare with observational resuls, we apply the model to the  massive 
(M$_1$+M$_2$=34.5 + 15.7 M$_\odot$, $P$=33\,d, $e$=0.5) heartbeat star MACHO 80.7443.1718 which displays 
an unusually large ($\sim$23\%) brightness increase at periastron.
Our model directly predicts a $\sim$23\% increase in tidal shear energy dissipation rates
at periastron.  It thus accounts in a straightforward manner for the extraordinarily large
brightness increase that is observed.  However, this  assumes that all of the dissipated energy
contributes to the luminosity.  If part of the energy remains within the star, the radiated
energy would be smaller.  On the other hand, however, the energy that may be trapped in sub-photospheric
layers could trigger a temporary increase in the stellar radius which would then lead to even larger 
tidal perturbations. Such processes would be interesting to analyze in the future.

Based on the results of our grid of models discussed in the first part of the paper, we are led 
to conclude that MACHO 80.7443.1718's unusually strong periastron effect is due to its highly 
super-synchronous rotation rate ($\beta_0 \sim$1.95), a state that is not expected to be very 
common in binary systems if they have had time to pseudosynchronize.  MACHO 80.7443.1718 
belongs to the  Lucke-Hodge 58 association which has an estimated age of only a couple of 
million years \citep{1994AJ....108.1256G}, which could explain why it still is in such rapid 
rotation.  It is also interesting to note that its $\beta_0$ value exceeds the limit 
$\beta_{crit}$=18/11 at which the eccentricity tends to increase over time instead of decrease 
\citep{2008EAS....29...67Z}.  

Our model for  MACHO 80.7443.1718 does not predict a dip in the brightness, whereas it's observed
light curve does show such a dip. \citet{2023NatAs...7.1218M} suggested that the dip could be due
to obscuration caused by ejected material.    It is tempting to speculate that the steep surface angular 
velocity gradients and fast bulge radius increase that take place around periastron could trigger 
brief mass-shedding events.  In this respect, a determination of the mass-loss rates at different 
orbital phases would be most useful.

The effect of tidal shear energy dissipation is often neglected when considering the structure 
of binary stars and their observational properties. This neglect is due to gaps in our 
understanding of the dissipation processes that are active and the assumption that viscous shear 
energy dissipation is negligible in stars.\footnote{This process, however, plays a major role 
in celestial objects such as Jupiter's moons Io and Europa and most likely play a role in 
close-in exoplanets.}  However, if it is non-negligible, as we assume in this paper, it can 
have a significant effect on the stellar properties which, in eccentric systems, become 
strongly orbital phase-dependent.  

\begin{acknowledgements} 

We thank Werner Schmutz for a critical reading of this paper and the anonymous referee who 
provided thoughtful comments and suggestions. GK is grateful to the Astronomy Department of the 
Indiana University for hosting a sabbatical visit during which the research in this paper was 
concluded. This work was supported by UNAM-PAPIIT 105723.

\end{acknowledgements}
\bibliographystyle{aa} 
\bibliography{TIDES_2023dec06.bib}        

\begin{thebibliography}{76}
\expandafter\ifx\csname natexlab\endcsname\relax\def\natexlab#1{#1}\fi

\bibitem[{{Alcock} {et~al.}(1997){Alcock}, {Allsman}, {Alves}, {Axelrod},
  {Becker}, {Bennett}, {Cook}, {Freeman}, {Griest}, {Guern}, {Lehner},
  {Marshall}, {Peterson}, {Pratt}, {Quinn}, {Rodgers}, {Stubbs}, {Sutherland},
  \& {Welch}}]{1997ApJ...486..697A}
{Alcock}, C., {Allsman}, R.~A., {Alves}, D., {et~al.} 1997, \apj, 486, 697

\bibitem[{{Astoul} {et~al.}(2021){Astoul}, {Park}, {Mathis}, {Baruteau}, \&
  {Gallet}}]{2021A&A...647A.144A}
{Astoul}, A., {Park}, J., {Mathis}, S., {Baruteau}, C., \& {Gallet}, F. 2021,
  \aap, 647, A144

\bibitem[{{Barker} \& {Astoul}(2021)}]{2021MNRAS.506L..69B}
{Barker}, A.~J. \& {Astoul}, A. A.~V. 2021, \mnras, 506, L69

\bibitem[{{Baruteau} \& {Rieutord}(2013)}]{2013JFM...719...47B}
{Baruteau}, C. \& {Rieutord}, M. 2013, Journal of Fluid Mechanics, 719, 47

\bibitem[{{Caleo} {et~al.}(2016){Caleo}, {Balbus}, \&
  {Tognelli}}]{2016MNRAS.460..338C}
{Caleo}, A., {Balbus}, S.~A., \& {Tognelli}, E. 2016, \mnras, 460, 338

\bibitem[{{Dewberry} \& {Wu}(2024)}]{2024MNRAS.527.2288D}
{Dewberry}, J.~W. \& {Wu}, S.~C. 2024, \mnras, 527, 2288

\bibitem[{{Efroimsky}(2018)}]{2018Icar..300..223E}
{Efroimsky}, M. 2018, \icarus, 300, 223

\bibitem[{{Eggleton} {et~al.}(1998){Eggleton}, {Kiseleva}, \&
  {Hut}}]{1998ApJ...499..853E}
{Eggleton}, P.~P., {Kiseleva}, L.~G., \& {Hut}, P. 1998, \apj, 499, 853

\bibitem[{{Elliott} {et~al.}(2000){Elliott}, {Miesch}, \&
  {Toomre}}]{2000ApJ...533..546E}
{Elliott}, J.~R., {Miesch}, M.~S., \& {Toomre}, J. 2000, \apj, 533, 546

\bibitem[{{Estrella-Trujillo} {et~al.}(2023){Estrella-Trujillo}, {Arthur},
  {Koenigsberger}, \& {Moreno}}]{2023A&A...670A..44E}
{Estrella-Trujillo}, D., {Arthur}, S.~J., {Koenigsberger}, G., \& {Moreno}, E.
  2023, \aap, 670, A44

\bibitem[{{Fellay} {et~al.}(2024){Fellay}, {Dupret}, \&
  {Rosu}}]{2024arXiv240102573F}
{Fellay}, L., {Dupret}, M.~A., \& {Rosu}, S. 2024, arXiv e-prints,
  arXiv:2401.02573

\bibitem[{{Fuller}(2017)}]{2017MNRAS.472.1538F}
{Fuller}, J. 2017, \mnras, 472, 1538

\bibitem[{{Garaud} \& {Kulenthirarajah}(2016)}]{2016ApJ...821...49G}
{Garaud}, P. \& {Kulenthirarajah}, L. 2016, \apj, 821, 49

\bibitem[{{Garmany} {et~al.}(1994){Garmany}, {Massey}, \&
  {Parker}}]{1994AJ....108.1256G}
{Garmany}, C.~D., {Massey}, P., \& {Parker}, J.~W. 1994, \aj, 108, 1256

\bibitem[{{Goldreich} \& {Nicholson}(1989)}]{1989ApJ...342.1079G}
{Goldreich}, P. \& {Nicholson}, P.~D. 1989, \apj, 342, 1079

\bibitem[{{Guo}(2020)}]{2020svos.conf..203G}
{Guo}, Z. 2020, in Stars and their Variability Observed from Space, ed.
  C.~{Neiner}, W.~W. {Weiss}, D.~{Baade}, R.~E. {Griffin}, C.~C. {Lovekin}, \&
  A.~F.~J. {Moffat}, 203--207

\bibitem[{{Guo} {et~al.}(2020){Guo}, {Shporer}, {Hambleton}, \&
  {Isaacson}}]{2020ApJ...888...95G}
{Guo}, Z., {Shporer}, A., {Hambleton}, K., \& {Isaacson}, H. 2020, \apj, 888,
  95

\bibitem[{{Hambleton} {et~al.}(2013){Hambleton}, {Degroote}, {Conroy},
  {Bloemen}, {Kurtz}, {Thompson}, {Fuller}, {Giammarco}, {Pablo}, \&
  {Pr{\v{s}}a}}]{2013EAS....64..285H}
{Hambleton}, K., {Degroote}, P., {Conroy}, K., {et~al.} 2013, in EAS
  Publications Series, Vol.~64, EAS Publications Series, ed. K.~{Pavlovski},
  A.~{Tkachenko}, \& G.~{Torres}, 285--294

\bibitem[{{Harrington} {et~al.}(2009){Harrington}, {Koenigsberger}, {Moreno},
  \& {Kuhn}}]{2009ApJ...704..813H}
{Harrington}, D., {Koenigsberger}, G., {Moreno}, E., \& {Kuhn}, J. 2009, \apj,
  704, 813

\bibitem[{{Harrington} {et~al.}(2016){Harrington}, {Koenigsberger},
  {Olgu{\'\i}n}, {Ilyin}, {Berdyugina}, {Lara}, \&
  {Moreno}}]{2016A&A...590A..54H}
{Harrington}, D., {Koenigsberger}, G., {Olgu{\'\i}n}, E., {et~al.} 2016, \aap,
  590, A54

\bibitem[{{Hirschi} \& {Maeder}(2010)}]{2010A&A...519A..16H}
{Hirschi}, R. \& {Maeder}, A. 2010, \aap, 519, A16

\bibitem[{{Hut}(1981)}]{1981A&A....99..126H}
{Hut}, P. 1981, \aap, 99, 126

\bibitem[{{Hutchings}(1979)}]{1979IAUS...83....3H}
{Hutchings}, J.~B. 1979, in Mass Loss and Evolution of O-Type Stars, ed. P.~S.
  {Conti} \& C.~W.~H. {De Loore}, Vol.~83, 3--20

\bibitem[{{Jayasinghe} {et~al.}(2021){Jayasinghe}, {Kochanek}, {Strader},
  {Stanek}, {Vallely}, {Thompson}, {Hinkle}, {Shappee}, {Dupree}, {Auchettl},
  {Chomiuk}, {Aydi}, {Dage}, {Hughes}, {Shishkovsky}, {Sokolovsky}, {Swihart},
  {Voggel}, \& {Thompson}}]{2021MNRAS.506.4083J}
{Jayasinghe}, T., {Kochanek}, C.~S., {Strader}, J., {et~al.} 2021, \mnras, 506,
  4083

\bibitem[{{Jayasinghe} {et~al.}(2020){Jayasinghe}, {Stanek}, {Kochanek},
  {Shappee}, {Holoien}, {Thompson}, {Prieto}, {Dong}, {Pawlak}, {Pejcha},
  {Shields}, {Pojmanski}, {Otero}, {Hurst}, {Britt}, \&
  {Will}}]{2020MNRAS.491...13J}
{Jayasinghe}, T., {Stanek}, K.~Z., {Kochanek}, C.~S., {et~al.} 2020, \mnras,
  491, 13

\bibitem[{{Jayasinghe} {et~al.}(2019){Jayasinghe}, {Stanek}, {Kochanek},
  {Thompson}, {Shappee}, \& {Fausnaugh}}]{2019MNRAS.489.4705J}
{Jayasinghe}, T., {Stanek}, K.~Z., {Kochanek}, C.~S., {et~al.} 2019, \mnras,
  489, 4705

\bibitem[{{K{\"a}pyl{\"a}} {et~al.}(2020){K{\"a}pyl{\"a}}, {Rheinhardt},
  {Brandenburg}, \& {K{\"a}pyl{\"a}}}]{2020A&A...636A..93K}
{K{\"a}pyl{\"a}}, P.~J., {Rheinhardt}, M., {Brandenburg}, A., \&
  {K{\"a}pyl{\"a}}, M.~J. 2020, \aap, 636, A93

\bibitem[{{Kirk} {et~al.}(2016){Kirk}, {Conroy}, {Pr{\v{s}}a}, {Abdul-Masih},
  {Kochoska}, {Matijevi{\v{c}}}, {Hambleton}, {Barclay}, {Bloemen}, {Boyajian},
  {Doyle}, {Fulton}, {Hoekstra}, {Jek}, {Kane}, {Kostov}, {Latham}, {Mazeh},
  {Orosz}, {Pepper}, {Quarles}, {Ragozzine}, {Shporer}, {Southworth},
  {Stassun}, {Thompson}, {Welsh}, {Agol}, {Derekas}, {Devor}, {Fischer},
  {Green}, {Gropp}, {Jacobs}, {Johnston}, {LaCourse}, {Saetre}, {Schwengeler},
  {Toczyski}, {Werner}, {Garrett}, {Gore}, {Martinez}, {Spitzer}, {Stevick},
  {Thomadis}, {Vrijmoet}, {Yenawine}, {Batalha}, \&
  {Borucki}}]{2016AJ....151...68K}
{Kirk}, B., {Conroy}, K., {Pr{\v{s}}a}, A., {et~al.} 2016, \aj, 151, 68

\bibitem[{{Koenigsberger} \& {Moreno}(2016)}]{2016RMxAA..52..113K}
{Koenigsberger}, G. \& {Moreno}, E. 2016, \rmxaa, 52, 113

\bibitem[{{Koenigsberger} {et~al.}(2021){Koenigsberger}, {Moreno}, \&
  {Langer}}]{2021A&A...653A.127K}
{Koenigsberger}, G., {Moreno}, E., \& {Langer}, N. 2021, \aap, 653, A127

\bibitem[{{Ko{\l}aczek-Szyma{\'n}ski}
  {et~al.}(2021){Ko{\l}aczek-Szyma{\'n}ski}, {Pigulski}, {Michalska},
  {Mo{\'z}dzierski}, \& {R{\'o}{\.z}a{\'n}ski}}]{2021A&A...647A..12K}
{Ko{\l}aczek-Szyma{\'n}ski}, P.~A., {Pigulski}, A., {Michalska}, G.,
  {Mo{\'z}dzierski}, D., \& {R{\'o}{\.z}a{\'n}ski}, T. 2021, \aap, 647, A12

\bibitem[{{Ko{\l}aczek-Szyma{\'n}ski}
  {et~al.}(2022){Ko{\l}aczek-Szyma{\'n}ski}, {Pigulski}, {Wrona}, {Ratajczak},
  \& {Udalski}}]{2022A&A...659A..47K}
{Ko{\l}aczek-Szyma{\'n}ski}, P.~A., {Pigulski}, A., {Wrona}, M., {Ratajczak},
  M., \& {Udalski}, A. 2022, \aap, 659, A47

\bibitem[{{Kumar} {et~al.}(1995){Kumar}, {Ao}, \&
  {Quataert}}]{1995ApJ...449..294K}
{Kumar}, P., {Ao}, C.~O., \& {Quataert}, E.~J. 1995, \apj, 449, 294

\bibitem[{{Lai}(1997)}]{1997ApJ...490..847L}
{Lai}, D. 1997, \apj, 490, 847

\bibitem[{{Landau} \& {Lifshitz}(1987)}]{1987flme.book.....L}
{Landau}, L.~D. \& {Lifshitz}, E.~M. 1987, {Fluid Mechanics}

\bibitem[{{Le Bars} {et~al.}(2022){Le Bars}, {Barik}, {Burmann}, {Lathrop},
  {Noir}, {Schaeffer}, \& {Triana}}]{2022SGeo...43..229L}
{Le Bars}, M., {Barik}, A., {Burmann}, F., {et~al.} 2022, Surveys in
  Geophysics, 43, 229

\bibitem[{{Lee}(2017)}]{2017MNRAS.468.1864L}
{Lee}, U. 2017, \mnras, 468, 1864

\bibitem[{{Lurie} {et~al.}(2017){Lurie}, {Vyhmeister}, {Hawley}, {Adilia},
  {Chen}, {Davenport}, {Juri{\'c}}, {Puig-Holzman}, \&
  {Weisenburger}}]{2017AJ....154..250L}
{Lurie}, J.~C., {Vyhmeister}, K., {Hawley}, S.~L., {et~al.} 2017, \aj, 154, 250

\bibitem[{{MacLeod} \& {Loeb}(2023)}]{2023NatAs...7.1218M}
{MacLeod}, M. \& {Loeb}, A. 2023, Nature Astronomy, 7, 1218

\bibitem[{{Maeder} \& {Meynet}(1996)}]{1996A&A...313..140M}
{Maeder}, A. \& {Meynet}, G. 1996, \aap, 313, 140

\bibitem[{{Maeder} \& {Meynet}(2000)}]{2000ARA&A..38..143M}
{Maeder}, A. \& {Meynet}, G. 2000, \araa, 38, 143

\bibitem[{{Mathis} {et~al.}(2016){Mathis}, {Auclair-Desrotour}, {Guenel},
  {Gallet}, \& {Le Poncin-Lafitte}}]{2016A&A...592A..33M}
{Mathis}, S., {Auclair-Desrotour}, P., {Guenel}, M., {Gallet}, F., \& {Le
  Poncin-Lafitte}, C. 2016, \aap, 592, A33

\bibitem[{{Mathis} {et~al.}(2004){Mathis}, {Palacios}, \&
  {Zahn}}]{2004A&A...425..243M}
{Mathis}, S., {Palacios}, A., \& {Zahn}, J.~P. 2004, \aap, 425, 243

\bibitem[{{McMillan} {et~al.}(1987){McMillan}, {McDermott}, \&
  {Taam}}]{1987ApJ...318..261M}
{McMillan}, S. L.~W., {McDermott}, P.~N., \& {Taam}, R.~E. 1987, \apj, 318, 261

\bibitem[{{Moreno} \& {Koenigsberger}(1999)}]{1999RMxAA..35..157M}
{Moreno}, E. \& {Koenigsberger}, G. 1999, \rmxaa, 35, 157

\bibitem[{Moreno {et~al.}(2011)Moreno, Koenigsberger, \&
  Harrington}]{Moreno:2011jq}
Moreno, E., Koenigsberger, G., \& Harrington, D.~M. 2011, A{\&}A, 528, 48

\bibitem[{Moreno {et~al.}(2005)Moreno, Koenigsberger, \&
  Toledano}]{Moreno:2005cq}
Moreno, E., Koenigsberger, G., \& Toledano, O. 2005, A{\&}A, 437, 641

\bibitem[{{Ogilvie}(2014)}]{2014ARA&A..52..171O}
{Ogilvie}, G.~I. 2014, \araa, 52, 171

\bibitem[{{Orosz} \& {Hauschildt}(2000)}]{2000A&A...364..265O}
{Orosz}, J.~A. \& {Hauschildt}, P.~H. 2000, \aap, 364, 265

\bibitem[{{Pablo} {et~al.}(2017){Pablo}, {Richardson}, {Fuller}, {Rowe},
  {Moffat}, {Kuschnig}, {Popowicz}, {Handler}, {Neiner}, {Pigulski}, {Wade},
  {Weiss}, {Buysschaert}, {Ramiaramanantsoa}, {Bratcher}, {Gerhartz}, {Greco},
  {Hardegree-Ullman}, {Lembryk}, \& {Oswald}}]{2017MNRAS.467.2494P}
{Pablo}, H., {Richardson}, N.~D., {Fuller}, J., {et~al.} 2017, \mnras, 467,
  2494

\bibitem[{{Palate} {et~al.}(2013{\natexlab{a}}){Palate}, {Koenigsberger},
  {Rauw}, {Harrington}, \& {Moreno}}]{2013A&A...556A..49P}
{Palate}, M., {Koenigsberger}, G., {Rauw}, G., {Harrington}, D., \& {Moreno},
  E. 2013{\natexlab{a}}, \aap, 556, A49

\bibitem[{{Palate} {et~al.}(2013{\natexlab{b}}){Palate}, {Rauw},
  {Koenigsberger}, \& {Moreno}}]{2013A&A...552A..39P}
{Palate}, M., {Rauw}, G., {Koenigsberger}, G., \& {Moreno}, E.
  2013{\natexlab{b}}, \aap, 552, A39

\bibitem[{{Penev} {et~al.}(2007){Penev}, {Sasselov}, {Robinson}, \&
  {Demarque}}]{2007ApJ...655.1166P}
{Penev}, K., {Sasselov}, D., {Robinson}, F., \& {Demarque}, P. 2007, \apj, 655,
  1166

\bibitem[{{P{\'e}rez-de-Tejada}(1999)}]{1999ApJ...525L..65P}
{P{\'e}rez-de-Tejada}, H. 1999, \apjl, 525, L65

\bibitem[{{Pigulski}(2018)}]{2018pas7.conf..151P}
{Pigulski}, A. 2018, in XXXVIII Polish Astronomical Society Meeting, ed.
  A.~{R{\'o}\&{\.z}a{\'n}ska}, Vol.~7, 151--156

\bibitem[{{Press} \& {Teukolsky}(1977)}]{1977ApJ...213..183P}
{Press}, W.~H. \& {Teukolsky}, S.~A. 1977, \apj, 213, 183

\bibitem[{{Richard} \& {Zahn}(1999)}]{1999A&A...347..734R}
{Richard}, D. \& {Zahn}, J.-P. 1999, \aap, 347, 734

\bibitem[{{Richardson} {et~al.}(2018){Richardson}, {Pablo}, {Sterken},
  {Pigulski}, {Koenigsberger}, {Moffat}, {Madura}, {Hamaguchi}, {Corcoran},
  {Damineli}, {Gull}, {Hillier}, {Weigelt}, {Handler}, {Popowicz}, {Wade},
  {Weiss}, \& {Zwintz}}]{2018MNRAS.475.5417R}
{Richardson}, N.~D., {Pablo}, H., {Sterken}, C., {et~al.} 2018, \mnras, 475,
  5417

\bibitem[{{Savonije} {et~al.}(1995){Savonije}, {Papaloizou}, \&
  {Alberts}}]{1995MNRAS.277..471S}
{Savonije}, G.~J., {Papaloizou}, J.~C.~B., \& {Alberts}, F. 1995, \mnras, 277,
  471

\bibitem[{{Scharlemann}(1981)}]{1981ApJ...246..292S}
{Scharlemann}, E.~T. 1981, \apj, 246, 292

\bibitem[{{Schenk} {et~al.}(2001){Schenk}, {Arras}, {Flanagan}, {Teukolsky}, \&
  {Wasserman}}]{2001PhRvD..65b4001S}
{Schenk}, A.~K., {Arras}, P., {Flanagan}, {\'E}.~{\'E}., {Teukolsky}, S.~A., \&
  {Wasserman}, I. 2001, \prd, 65, 024001

\bibitem[{{Shi} {et~al.}(2006){Shi}, {Wen}, {Bao}, {Li}, \&
  {Huang}}]{2006ScChD..49..492S}
{Shi}, X., {Wen}, D., {Bao}, X., {Li}, C., \& {Huang}, Y. 2006, Science China
  Earth Sciences, 49, 492

\bibitem[{{Sterken} \& {Breysacher}(1997)}]{1997A&A...328..269S}
{Sterken}, C. \& {Breysacher}, J. 1997, \aap, 328, 269

\bibitem[{{Sun} {et~al.}(2023){Sun}, {Townsend}, \&
  {Guo}}]{2023ApJ...945...43S}
{Sun}, M., {Townsend}, R.~H.~D., \& {Guo}, Z. 2023, \apj, 945, 43

\bibitem[{{Symon}(1971)}]{1971Symon}
{Symon}, K.~R. 1971, Mechanics (Addison-Wesley)

\bibitem[{{Talon} {et~al.}(1997){Talon}, {Zahn}, {Maeder}, \&
  {Meynet}}]{1997A&A...322..209T}
{Talon}, S., {Zahn}, J.~P., {Maeder}, A., \& {Meynet}, G. 1997, \aap, 322, 209

\bibitem[{{Terquem}(2021)}]{2021MNRAS.503.5789T}
{Terquem}, C. 2021, \mnras, 503, 5789

\bibitem[{{Thompson} {et~al.}(2012){Thompson}, {Everett}, {Mullally},
  {Barclay}, {Howell}, {Still}, {Rowe}, {Christiansen}, {Kurtz}, {Hambleton},
  {Twicken}, {Ibrahim}, \& {Clarke}}]{2012ApJ...753...86T}
{Thompson}, S.~E., {Everett}, M., {Mullally}, F., {et~al.} 2012, \apj, 753, 86

\bibitem[{{Tkachenko} {et~al.}(2016){Tkachenko}, {Matthews}, {Aerts},
  {Pavlovski}, {P{\'a}pics}, {Zwintz}, {Cameron}, {Walker}, {Kuschnig},
  {Degroote}, {Debosscher}, {Moravveji}, {Kolbas}, {Guenther}, {Moffat},
  {Rowe}, {Rucinski}, {Sasselov}, \& {Weiss}}]{2016MNRAS.458.1964T}
{Tkachenko}, A., {Matthews}, J.~M., {Aerts}, C., {et~al.} 2016, \mnras, 458,
  1964

\bibitem[{{Toledano} {et~al.}(2007){Toledano}, {Moreno}, {Koenigsberger},
  {Detmers}, \& {Langer}}]{2007A&A...461.1057T}
{Toledano}, O., {Moreno}, E., {Koenigsberger}, G., {Detmers}, R., \& {Langer},
  N. 2007, \aap, 461, 1057

\bibitem[{{Townsend} \& {Sun}(2023)}]{2023ApJ...953...48T}
{Townsend}, R.~H.~D. \& {Sun}, M. 2023, \apj, 953, 48

\bibitem[{{van Genderen} \& {Sterken}(2007)}]{2007IBVS.5782....1V}
{van Genderen}, A.~M. \& {Sterken}, C. 2007, Information Bulletin on Variable
  Stars, 5782, 1

\bibitem[{{Zahn}(1975)}]{1975A&A....41..329Z}
{Zahn}, J.~P. 1975, \aap, 41, 329

\bibitem[{{Zahn}(1977)}]{1977A&A....57..383Z}
{Zahn}, J.~P. 1977, \aap, 57, 383

\bibitem[{{Zahn}(1992)}]{1992A&A...265..115Z}
{Zahn}, J.~P. 1992, \aap, 265, 115

\bibitem[{{Zahn}(2008)}]{2008EAS....29...67Z}
{Zahn}, J.~P. 2008, in EAS Publications Series, Vol.~29, EAS Publications
  Series, ed. M.~J. {Goupil} \& J.~P. {Zahn}, 67--90

\end{thebibliography}

\clearpage
\vfill\eject

\begin{appendix}

\section{Extremum value of the angular velocity at each timestep}

\begin{figure}
\centering
\includegraphics[width=0.48\columnwidth]{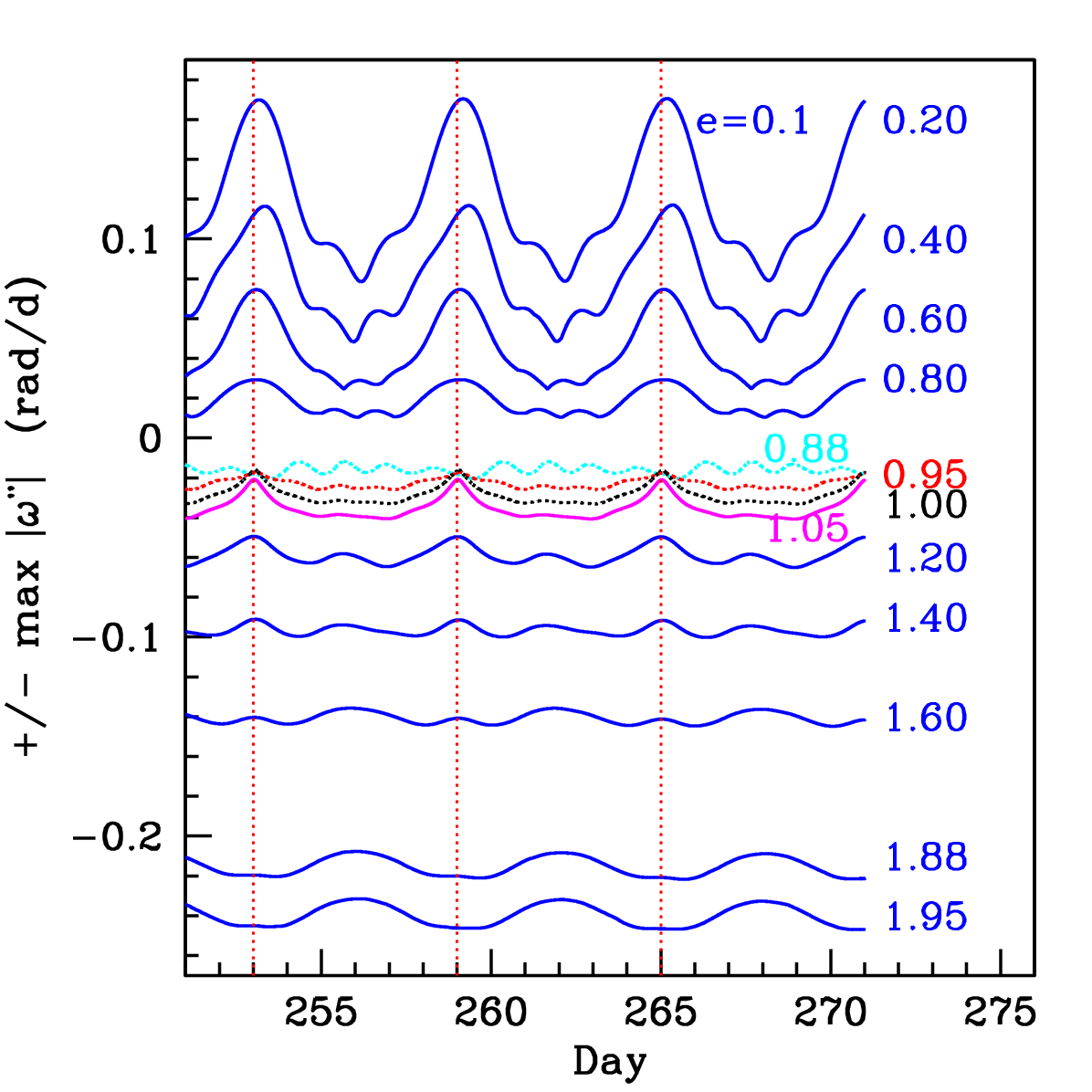}
\includegraphics[width=0.48\columnwidth]{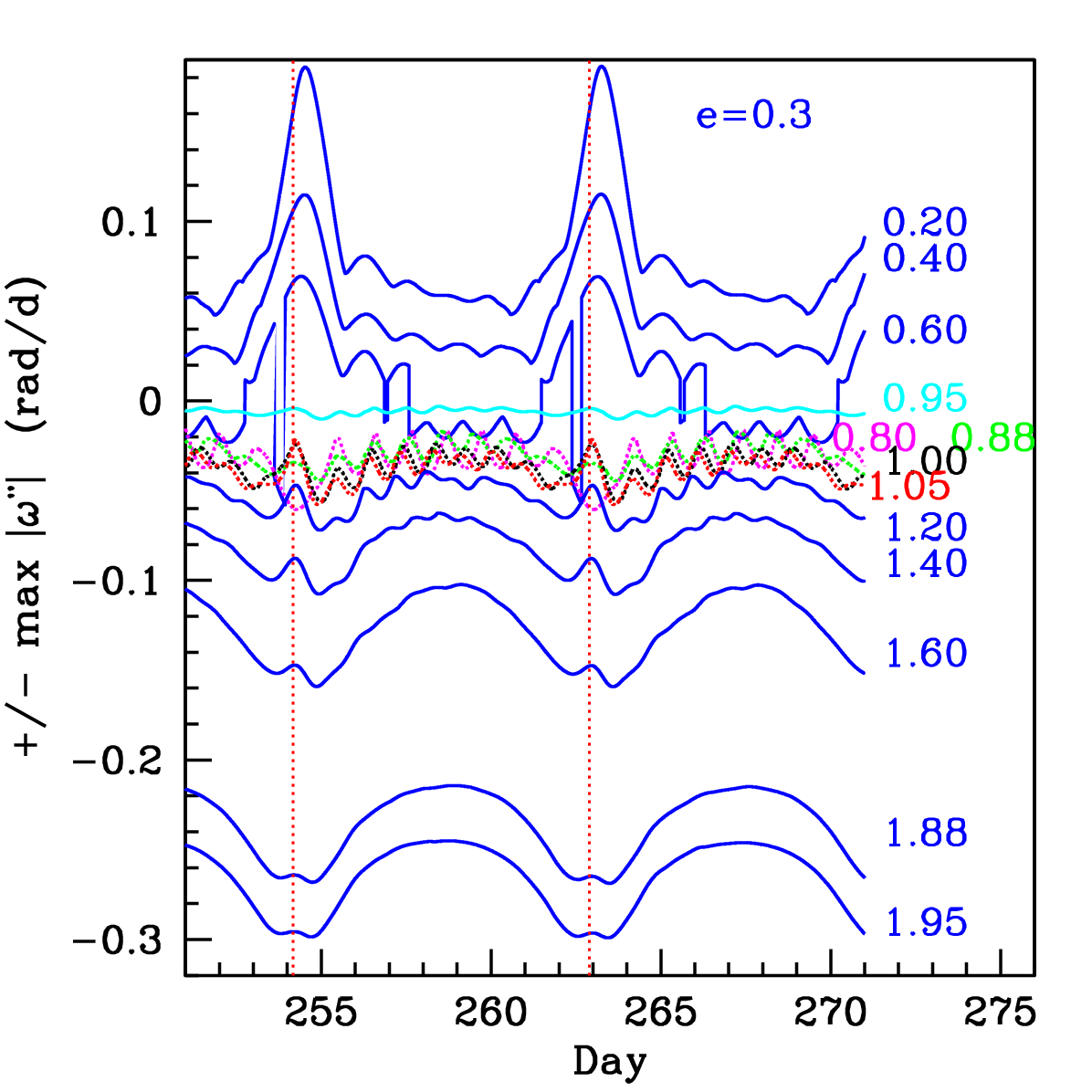}
\caption{Maximum perturbation of the surface angular velocity at the equator plotted as a function
of time for $e$=0.1 (left) and $e$=0.3 (right).  The velocity is measured in the rest reference
frame of the rotating star. Positive values indicate motion in the direction  of the orbital motion,
negative values to motion opposite to the orbital motion. Vertical lines indicate times of periastron.
Each curve is labeled with its value of $\beta_0$. The smallest tidal perturbations for $e$=0.1  occur
for $\beta_0$=0.88. For  $e$=0.3,  the $\beta_0$=0.80, 0.88, 1.00 and 1.05 curves show low amplitude
oscillations and  are very similar.  The smallest perturbations are observed for $\beta_0$=0.95.
Observational perspective: Surface layer is rotating faster than the inner core for $\beta_0<$1 and rotating
slower in $\beta_0>$1 models. Also, these stars are likely to display variations in the shape of the
photospheric line profiles as discussed in Moreno et al. (2005) and Harrington et al. (2009, 2016).
TEOs are evident mostly for $\beta_0 \sim$1 models.
\label{fig_omdp_e0.1and0.3}}
\end{figure}

\begin{figure}
\centering
\includegraphics[width=0.48\columnwidth]{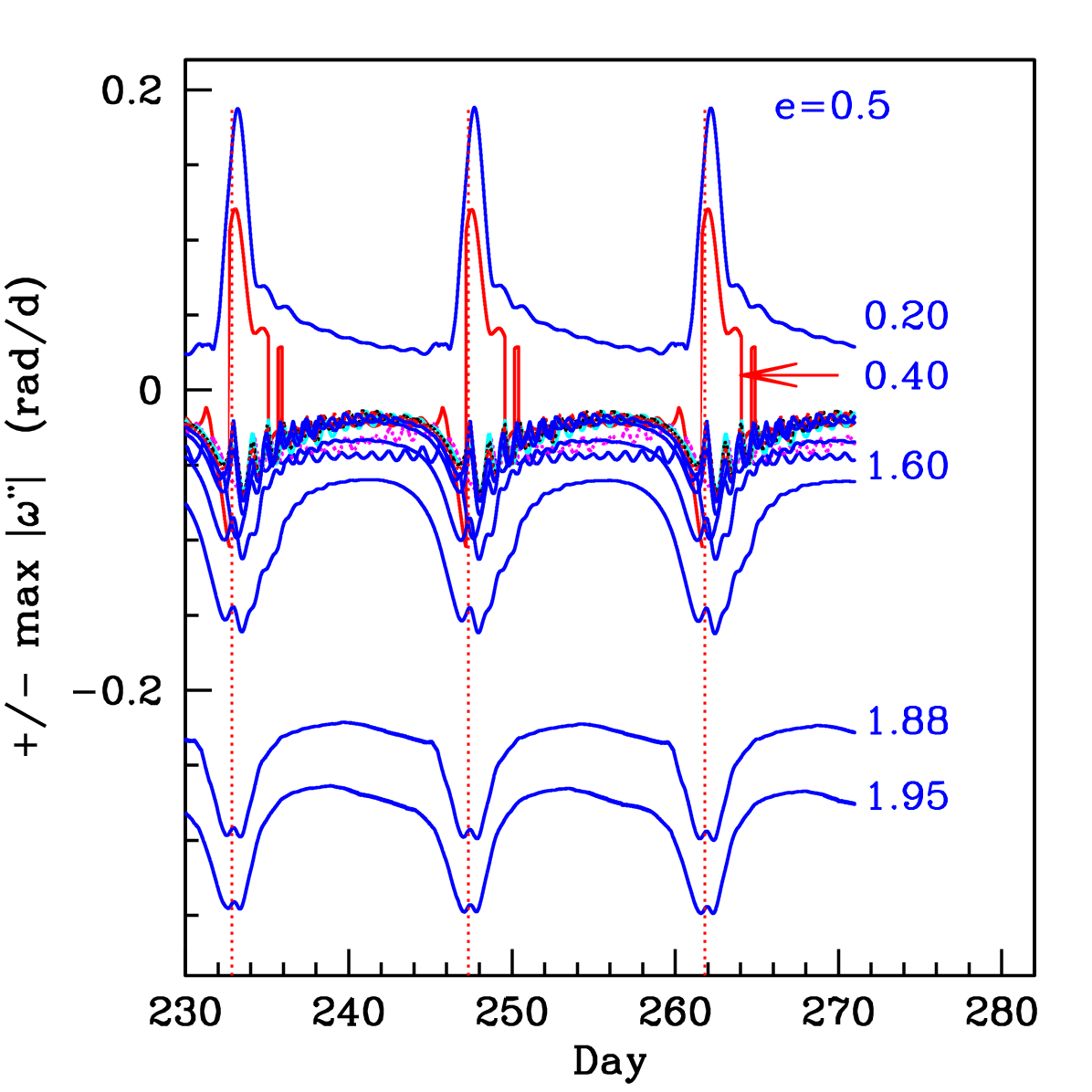}
\includegraphics[width=0.48\columnwidth]{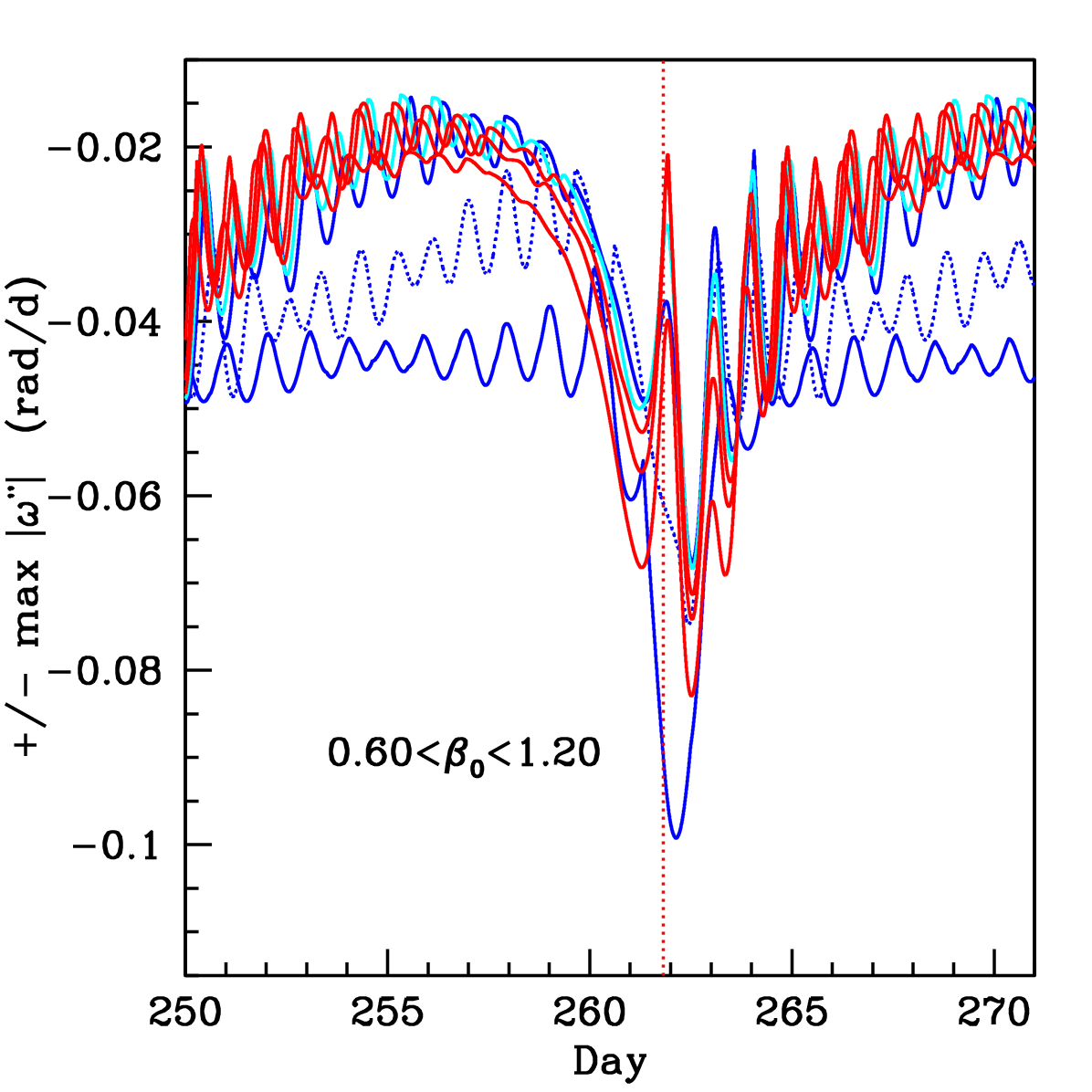}
\caption{Maximum  perturbation of the surface angular velocity at the equator for $e$=0.5 plotted
as a function of time.  The velocity is measured in the reference frame at rest in the rotating star. 
Positive values indicate motion in the direction  of the orbital motion, negative values motion 
opposite to the orbital motion. Vertical lines indicate times of periastron. {\it Left:} All models with
$\beta_0$ increasing from top to bottom. Models close to pseudosynchronism are not labeled 
for clarity in the figure.  {\it Right:} Models with 0.60$\leq\beta_0 \leq$1.00 are plotted in
blue and cyan and those with 1.05$\leq\beta_0 \leq$1.20 in red.                         
\label{fig_omdp_e0.5}}
\end{figure}

The eccentric binary models all display a variability pattern in the tidal perturbations  on the 
orbital timescale, the largest amplitudes occurring around the time of periastron. We quantify 
the time-dependent perturbations at each computational timestep by finding the largest deviation 
of the angular velocity from the underlying rigidly rotating core at each computational time step.
We define this quantity as max(|$\omega(r, \theta, \varphi, t|$),
where $\omega$ is the instantaneous angular velocity of the volume element located at
($r, \theta,\varphi$) at time $t$ as measured in the reference frame that rotates with the
primary star. Once this quantity is located at each timestep of the computation, it is assigned a positive
or negative sign depending on whether this maximum velocity is in the direction of the orbital motion
or in the opposite direction.  The first case occurs mostly in the subsynchronous models ($\beta_0<$1)
where the tidal force acts to increase the velocity toward corotation. The second case occurs in the
supersynchronous models where the tidal force acts to slow the velocity down.  This general behavior
is illustrated in Fig.~\ref{fig_omdp_e0.1and0.3} for  $e$=0.1 and 0.3 models and Fig.~\ref{fig_omdp_e0.5} 
for $e$=0.5. 

Around the time of periastron, the perturbation becomes strongest and the angular velocity changes
rapidly creating either a maximum (subsynchronous systems) or a minimum peak 
(highly supersynchronous systems).  Systems with rotation speeds that are not very far from corotation
at periastron tend to present high frequency oscillations throughout the orbital cycle and a variety
of behaviors at periastron.

Some of the models show a maximum peak at periastron and negative values at other phases  (for
example, $e$=0.3, $\beta_0$=0.60 and $e$=0.5, $\beta_0$=0.40) but in general have either positive or
negative values throughout the orbit.

\begin{table*}[t]
\label{tab:table7}
\centering
\caption{Turbulent viscosity ten layer models. \label{table_viscosities}}
\begin{tabular}{cccccccccccccc}
\hline
\hline
Model & $\beta_0$& V$_{rot}$& $<\dot{E}_{tot}>$ &\multicolumn{10}{c}{$\log\left< \nu/R_\odot^2/d \right>$} \\ 
      &            &           &                 &\multicolumn{5}{c}{Periastron}&\multicolumn{5}{c}{Apastron} \\
\hline
\multicolumn{4}{c}{}&k=2&k=4&k=6&k=8&k=10 & k=2&k=4&k=6&k=8&k=10\\
\hline
\multicolumn{14}{c}{$e$=0.1 ~~~~~~~~~ P=6 d  ~~~~~~~~~~~~}\\
\hline
201&0.20&11&7.61e-1&-3.20&-2.86&-2.72&-2.29&-1.17&-3.20&-2.89&-2.72&-2.78&-1.96\\
202&0.40&22&2.26e-1&-3.60&-3.03&-2.86&-2.43&-1.36&-3.59&-3.03&-2.84&-2.97&-1.94\\
203&0.60&34&4.78e-2&-3.81&-3.33&-3.02&-2.62&-1.50&-4.18&-3.32&-3.02&-3.12&-2.28\\
204&0.80&45&4.19e-3&-3.80&-3.95&-3.64&-2.92&-1.81&-4.59&-4.20&-3.64&-3.37&-1.97\\
205&0.88&49&2.48e-3&-3.87&-4.35&-3.66&-3.12&-1.91&-4.76&-4.50&-3.81&-3.21&-1.87\\
206&0.95&53&4.19e-3&-3.84&-4.04&-3.36&-3.33&-2.10&-4.74&-4.06&-3.36&-3.09&-1.83\\
207&1.00&56&8.78e-3&-3.85&-3.78&-3.21&-3.26&-2.21&-4.85&-3.78&-3.21&-2.99&-1.78\\
208&1.05&59&1.48e-2&-3.86&-3.56&-3.12&-3.17&-2.35&-4.61&-3.55&-3.12&-2.93&-1.73\\
209&1.20&67&6.32e-2&-3.86&-3.19&-2.94&-2.96&-1.96&-4.10&-3.19&-2.94&-2.79&.1.61\\
210&1.40&79&2.69e-1&-3.44&-2.95&-2.78&-2.67&-1.61&-3.43&-2.95&-2.78&-2.64&-1.47\\
211&1.60&90&8.33e-1&-3.12&-2.81&-2.68&-2.49&-1.40&-3.11&-2.81&-2.68&-2.52&-1.36\\
212&1.88&105&3.01&-2.88&-2.66&-2.56&-2.29&-1.18&-2.88&-2.66&-2.57&-2.36&-1.18\\
213&1.95&109&4.20  &-2.83&-2.63&-2.54&-2.25&-1.12&-2.83&-2.63&-2.54&-2.32&-1.13\\
\hline
\multicolumn{14}{c}{$e$=0.3 ~~~~~~~~~ P=8.73\,d  ~~~~~~~~~~~~}\\
\hline
   &    &   &      &&&&&\\
301&0.20&13&6.61e-1&-3.55&-3.02&-2.83&-2.14&-1.09&-3.54&-3.02&-2.85&-2.77&-1.61\\
302&0.40&27&1.81e-1&-4.16&-3.42&-3.12&-2.29&-1.27&-4.15&-3.42&-3.12&-3.14&-1.70\\
303&0.60&40&5.28e-2&-4.22&-3.85&-3.35&-2.46&-1.35&-4.40&-4.14&-3.73&-3.24&-1.63\\
304&0.80&53&2.47e-2&-4.29&-3.88&-3.26&-2.66&-1.51&-4.40&-3.88&-3.26&-3.29&-1.49\\
305&0.88&59&1.58e-2&-4.46&-3.88&-3.30&-2.90&-1.75&-4.54&-3.86&-3.30&-3.23&-1.40\\
306&0.95&63&4.21e-4&-4.37&-4.42&-4.00&-3.92&-2.40&-4.55&-4.21&-3.73&-3.81&-1.30\\
307&1.00&67&2.34e-2&-4.47&-3.61&-3.18&-3.32&-2.02&-4.44&-3.60&-3.18&-3.03&-1.32\\
308&1.05&70&2.66e-2&-4.48&-3.56&-3.16&-3.43&-2.13&-4.27&-3.55&-3.16&-2.98&-1.29\\
309&1.20&80&5.80e-2&-4.25&-3.32&-3.03&-2.97&-1.86&-4.26&-3.32&-3.03&-2.96&-1.22\\
310&1.40&93&2.02e-1&-3.74&-3.06&-2.86&-2.62&-1.53&-3.72&-3.06&-2.86&-2.84&-1.14\\
311&1.60&107&7.18e-1&-3.21&-2.85&-2.70&-2.41&-1.31&-3.20&-2.85&-2.70&-2.69&-1.11\\
312&1.88&125&4.36  &-2.80&-2.60&-2.49&-2.21&-1.09&-2.80&-2.60&-2.49&-2.52&-1.14\\
313&1.95&130&6.45  &-2.75&-2.56&-2.46&-2.17&-1.03&-2.75&-2.56&-2.46&-2.44&-1.10\\
\hline
\multicolumn{9}{c}{$e$=0.0 P=6\,d}& \multicolumn{4}{c}{}      \\
\hline
102&0.40&18&1.83e-1&-3.23&-3.04&-2.96&-2.74&-1.62  &       &       &       &        &    \\
103&0.60&27&3.79e-2&-4.20&-3.29&-3.00&-2.88&-1.79  &       &       &       &        &   \\
104&0.80&36&6.01e-3&-4.09&-3.53&-3.34&-3.22&-2.10  &       &       &       &        &   \\
106&0.95&43&2.09e-4&-4.28&-4.31&-3.96&-3.76&-2.70  &       &       &       &        &   \\
108&1.05&48&1.41e-3&-4.65&-4.43&-3.93&-3.80&-2.70   &       &       &       &        &   \\
109&1.20&55&4.13e-3&-4.22&-3.86&-3.30&-3.16&-2.10   &       &       &       &        &   \\
111&1.60&73&1.52e-1&-3.67&-3.04&-2.84&-2.73&-1.61   &       &       &       &        &    \\
113&1.95&89&9.27e-1&-3.12&-2.81&-2.68&-2.51&-1.38    &       &       &       &        &   \\
\hline \hline
\end{tabular}
\tablefoot{V$_{rot}$ is the initial equatorial rotation velocity in units of km/s; $\dot{E}_{tot}$ 
is the total energy dissipation rate after the numerical simulation has evolved to 50 orbital 
periods in units of 10$^{35}$ ergs/s; $\nu$ is the viscosity as computed in the model using
Eq. 2 with $\lambda$=1 in units of R$_{\odot}^2$/d. $k$ indicates the layer with $k$=1 the layer
adjoining the core and $k$=10 the surface layer, each layer having a depth of 
$\Delta$R=0.329\,R$_\odot$. The header of each set of model grids indicates the corresponding
orbital period and eccentricity.}
\end{table*}

\end{appendix}
\end{document}